\documentstyle[11pt,aaspp4,flushrt]{article}

\def\ebox{$\sqcup$\llap{$\sqcap$}}      
\def\4c{4C~41.17}

\def\spose#1{\hbox to 0pt{#1\hss}}

\def\kms{\ifmmode {\rm\,km\,s^{-1}}\else
    ${\rm\,km\,s^{-1}}$\fi}
\def\kmsMpc{\ifmmode {\rm\,km\,s^{-1}\,Mpc^{-1}}\else
    ${\rm\,km\,s^{-1}\,Mpc^{-1}}$\fi}

\def\msun{\ifmmode {\rm\,M_\odot}\else ${\rm\,M_\odot}$\fi}
\def\Msun{\ifmmode {\rm\,M_\odot}\else ${\rm\,M_\odot}$\fi}
\def\lsun{\ifmmode {\rm\,L_\odot}\else ${\rm\,L_\odot}$\fi}
\def\Lsun{\ifmmode {\rm\,L_\odot}\else ${\rm\,L_\odot}$\fi}
\def\rsun{\ifmmode {\rm\,R_\odot}\else ${\rm\,R_\odot}$\fi}
\def\Rsun{\ifmmode {\rm\,R_\odot}\else ${\rm\,R_\odot}$\fi}

\def\cm{{\rm\,cm}}
\def\cm3{\ifmmode {\rm\,cm^{-3}}\else ${\rm\,cm^{-3}}$\fi}

\def\ergps{\ifmmode {\rm\,erg\,s^{-1}}\else ${\rm\,erg\,s^{-1}}$\fi}
\def\ergpscm2{\ifmmode {\rm\,erg\,s^{-1}\,cm^{-2}}\else
    ${\rm\,erg\,s^{-1}\,cm^{-2}}$\fi}

\def\eg{{e.g.}}
\def\deg{\ifmmode {^{\circ}}\else {$^\circ$}\fi}
\def\degr{\ifmmode {^{\circ}}\else {$^\circ$}\fi}
\def\degs{\ifmmode {^{\circ}}\else {$^\circ$}\fi}

\def\etal{{et al.~}}

\def\h3Mpc{h^{-3}{\rm Mpc}^3}
\def\Ho{\ifmmode {\rm\,H_0}\else ${\rm\,H_0}$\fi}
\def\hnot{\ifmmode {\rm\,H_0}\else ${\rm\,H_0}$\fi}
\def\h0{\ifmmode {\rm\,H_0}\else ${\rm\,H_0}$\fi}
\def\hnotunit{\ifmmode {\rm\,km\,s^{-1}\,Mpc^{-1}}\else
    ${\rm\,km\,s^{-1}\,Mpc^{-1}}$\fi}
\def\qnot{\ifmmode {\rm\,q_0}\else ${\rm q_0}$\fi}
\def\q0{\ifmmode {\rm\,q_0}\else ${\rm q_0}$\fi}
\def\ie{{i.e.}}
\def\mic{\ifmmode {\rm\,\mu m}\else ${\rm \mu m}$\fi}

\def\arcsec{\ifmmode {^{\prime\prime}}\else $^{\prime\prime}$\fi}
\def\asec{\ifmmode {^{\prime\prime}}\else $^{\prime\prime}$\fi}
\def\arcmin{\ifmmode {^{\prime}}\else $^{\prime}$\fi}
\def\amin{\ifmmode {^{\prime}}\else $^{\prime}$\fi}

\def\secper{\ifmmode \rlap.{^{s}}\else $\rlap{.}{^{s}} $\fi}
\def\minper{\ifmmode \rlap.{^{m}}\else $\rlap{.}{^m} $\fi}
\def\magper{\ifmmode \rlap.{^{m}}\else $\rlap{.}{^m} $\fi}
\def\farcs{\ifmmode \rlap.{^{\prime\prime}}\else
    $\rlap.{^{\prime\prime}}$\fi}
\def\arcsper{\ifmmode \rlap.{^{\prime\prime}}\else
    $\rlap.{^{\prime\prime}}$\fi}
\def\arcmper{\ifmmode \rlap.{^{\prime}}\else
    $\rlap.{^{\prime}}$\fi}
\def\spose#1{\hbox to 0pt{#1\hss}}
\def\simlt{\mathrel{\spose{\lower 3pt\hbox{$\mathchar"218$}}
     \raise 2.0pt\hbox{$\mathchar"13C$}}}
\def\simgt{\mathrel{\spose{\lower 3pt\hbox{$\mathchar"218$}}
     \raise 2.0pt\hbox{$\mathchar"13E$}}}

\def\refindent{\par\noindent\parskip=2pt\hangindent=3pc\hangafter=1 }

\def\araa{{ARA\&A}}
\def\aa{{A\&A}}

\def\aasupp{{A\&AS}}
\def\aj{{AJ}}
\def\apj{{ApJ}}
\def\apjlett{{ApJ}}

\def\apjsupp{{ApJS}}

\def\mn{{MNRAS}}
\def\mnras{{MNRAS}}
\def\nature{{Nature}}

\def\pasp{{PASP}}

\def\apjref#1;#2;#3;#4 {\par\pp#1, {#2}, #3, #4 \par}

\begin{document}

\title{Triggered Star Formation in a Massive Galaxy at z=3.8: 4C~41.17\altaffilmark{1}}
\author{Arjun Dey}
\affil{KPNO/NOAO\altaffilmark{2}, 950 N. Cherry Ave., P. O. Box 26732, Tucson, 
AZ 85726}
\affil{dey@noao.edu}
\author{Wil van Breugel}
\affil{Institute of Geophysics \& Planetary Physics, LLNL, Livermore, CA 94550}
\affil{wil@igpp.llnl.gov}
\author{William D.\ Vacca}
\affil{Institute for Astronomy, University of Hawaii, 2680 Woodlawn Dr., Honolulu, HI 96822}
\affil{vacca@athena.ifa.hawaii.edu}
\author{Robert Antonucci}
\affil{Physics Dept., University of California, Santa Barbara, CA 93106}
\affil{ski@chester.physics.ucsb.edu}

\vspace{2in}

\centerline {Accepted for publication in the December 1, 1997 Apj}
\altaffiltext{1}{Based on observations at the W.\ M.\ Keck Observatory.}
\altaffiltext{2}{The National Optical Astronomy Observatories are
operated by the Association of Universities for Research in Astronomy under
Cooperative Agreement with the National Science Foundation.}

\vfill\eject

\begin{abstract}

We present deep spectropolarimetric observations obtained with the
W.\ M.\ Keck Telescope of the very high redshift
($z$=3.79786$\pm$0.0024) radio galaxy \4c. We find that the bright,
spatially extended rest-frame UV continuum emission from this galaxy,
which is aligned with the radio axis, is unpolarized ($P_{2\sigma} <
2.4\%$).  This implies that scattered AGN light, which is generally the
dominant contributor to the rest-frame UV emission in $z\sim 1$ radio
galaxies, is unlikely to be a major component of the UV flux from \4c.
The resulting total light spectrum shows absorption lines and
P-Cygni--like features that are similar to those detected in the
spectra of the recently discovered population of star forming galaxies
at slightly lower ($z\sim2-3$) redshifts. It may be possible for a
galaxian outflow to contribute partially to the absorption line
profiles of the low ionization species; however, it is unlikely that
the high velocity wings of the high ionization lines are dominated by a
galaxian wind since the outflow mass implied by the absorption line
strengths is very large.  The detection of the \ion{S}{5}$\lambda$1502
stellar photospheric absorption line, the shape of the blue wing of the
\ion{Si}{4} profile, the unpolarized continuum emission, the inability
of any AGN-related processes to account for the UV continuum flux, and
the overall similarity of the UV continuum spectra of \4c\ and the
nearby star-forming region NGC~1741B1 strongly suggest that the UV
light from \4c\ is dominated by young, hot stars.  If all of the UV
emission is due to starlight from a young population, the implied
star-formation rate is roughly 140 -- 1100~$h_{50}^{-2}$~\Msun/yr.  The
deep spectroscopy presented here combined with the morphology of the
system at radio and optical wavelengths and the possibly comparable
ages for the radio source structure and the UV stellar population
suggest that star formation in \4c\ was triggered by the expansion of
the radio source into the ambient medium.  Our current observations are
consistent with the hypothesis that \4c\ is undergoing its major epoch
of star formation at $z\sim 4$, and that by $z\sim 1$ it will have
evolved to have spectral and morphological properties similar to those
observed in known $z\sim1$ powerful radio galaxies.

\end{abstract}


\vfill\eject

\section {Introduction}

Until recently, radio source samples provided astronomers with the only
known galaxy--like objects at cosmological distances. Currently, three
of the four brightest known visible galaxies at redshifts $z\simgt 4$
are radio galaxies. One of the most intriguing properties of powerful
radio galaxies at high-redshift is that their spatially extended UV
morphology is preferentially aligned with the major axis of their radio
emission (McCarthy \etal 1987, Chambers \etal 1987).  Several possible
explanations have been proposed for this `alignment effect' (see
McCarthy 1993 for a recent review); the two most compelling of these
are (1) that the UV morphology is dominated by dust- and
electron-scattered light from an anisotropically radiating nuclear
source (e.g., Tadhunter \etal 1988), and (2) that the UV morphology is
the result of star-formation triggered by the radio source as it
expands into the dense ambient medium (e.g., De Young 1981, 1989; Rees
1989; Begelman \& Cioffi 1989). Although there is circumstantial
evidence at low redshift for radio sources triggering
star-formation (e.g., in Minkowski's Object, van Breugel \etal 1985;
in 3C~285, van Breugel \& Dey 1993; and in the Abell 1795 cD galaxy, 
McNamara \etal 1996), the majority of powerful
$z\sim 1$ radio galaxies investigated exhibit strong polarization
($\sim 10\%$) of the continuum emission and broad emission lines, a
result which confirms that much of the spatially extended rest-frame UV
continuum emission in these systems is scattered light from a hidden
AGN (di Serego Alighieri \etal 1989, Tadhunter \etal 1988, Jannuzi \&
Elston 1991, Jannuzi \etal 1995, Dey \etal 1996, Cimatti \etal 1996,
1997). The fractional polarization in most of the $z\sim 1-2$ powerful
radio galaxies investigated thus far show either no significant
variation of polarization with wavelength (\ie,
$P(\lambda)\approx{\rm\ constant}$) or, in a few cases, polarization
increasing towards shorter wavelengths.  In addition, $z\sim 1$ radio
galaxies of lower radio power (\ie, in which the AGN makes an
insignificant contribution to the UV light) show underlying stellar
populations that are both dynamically and spectrally old (\eg, Spinrad
\etal 1997, Dunlop \etal 1996, Rigler \etal 1992), providing further
evidence that triggered star-formation is not the dominant process
responsible for the alignment effect in $z\sim 1$ radio galaxies.

Despite the success of the scattering hypothesis in explaining the
alignment effect in most $z\sim 1$ radio galaxies, the processes
responsible for its origin in the highest redshift systems remain open
to debate. The $z\simgt 3$ radio galaxies are faint, and until now have
remained beyond the reach of most modern polarimeters.

In this paper, we present spectropolarimetric observations of the
$z=3.80$ radio galaxy \4c. The broad-band spectral energy distribution
of \4c\  was first investigated by Chambers \etal (1990), who suggested
that starlight from a young population produced the bulk of the
observed optical and near-IR continuum emission, and that the alignment
of the UV morphology with the radio source axis is most likely due to
star-formation triggered by the radio source.  Although there is close
spatial coincidence between the compact UV and radio continuum emitting
components (Miley \etal 1992), the observed UV emission is vastly in
excess of that which can be produced by synchrotron emission from
relativistic electrons responsible for the radio emission, or inverse
Compton scattering of the radio or cosmic microwave background photons
off these electrons (Carilli \etal 1994).  Our data demonstrate that
the UV light in this system is not dominated by scattered light from an
active nucleus, but instead is very likely to be dominated by
starlight from a young stellar population.  We discuss our observations
of this high-redshift galaxy in the context of a scenario where the
$z\sim 4$ powerful radio galaxies are the progenitors of the $z\sim 1$
population.

Throughout this paper we assume that \hnot=50\hnotunit\ and \qnot=0.1.
The scale at $z=3.80$ is then 10.8$h_{50}^{-1}$~kpc/\arcsec\ and the
lookback time is nearly 14.3$h_{50}^{-1}$~Gyr, or 86\% of the age of
the universe.  In comparison, for \qnot=0.5 the angular scale is
6.6$h_{50}^{-1}$~kpc/\arcsec\ and the lookback time is
11.8$h_{50}^{-1}$~Gyr, or 90\% of the age of the universe.

\section {Observations}

We observed \4c\ on the nights of U.T.  1996 December 9 and 10 using
the spectropolarimetric mode of the Low Resolution Imaging Spectrometer
(Oke \etal 1995, Goodrich \etal 1995) at the Cassegrain focus of the
Keck II Telescope.  The observations were obtained using the 400
line/mm grating ($\lambda_{\rm blaze} = 8500$\AA; dispersion =
1.86\AA/pixel) to sample the wavelength range $\lambda\lambda$5500--9280\AA\ 
(corresponding to $\lambda\lambda$1146--1933\AA\ in the rest frame of
\4c), and a 1\arcsec\ wide slit (effective resolution FWHM in the
observed frame of $\approx$8\AA). The LRIS detector is a Tek 2048$^2$
CCD with 24$\mu$m pixels with a pixel scale of 0{\farcs}214
pix$^{-1}$.  The CCD read noise during our run was 6.2$e^-$ and the
gain was $\approx 2.1 e^-$/ADU.

Over the two nights, we obtained a total of seven sets of observations
(three on Dec.~9 and four on Dec.~10) with the spectrograph slit
oriented along the major axis of the rest-frame UV continuum emission
from the galaxy (PA=76.5$^\circ$). Each set is comprised of
observations made in four waveplate positions ($0^\circ$, $45^\circ$,
$22.5^\circ$, $67.5^\circ$) which sample the electric vector in four
position angles on the sky ($76.5^\circ$, $166.5^\circ$, $121.5^\circ$,
$31.5^\circ$ respectively). The exposure time per waveplate position
was 20~minutes for each set, resulting in a total exposure time of 
140~minutes per waveplate position, and 9.33 hours for the total spectrum.
After two sets of observations, we reacquired the galaxy using an offset from a
nearby star.  Our first night was affected by minor telescope focus problems
and poor seeing, but the seeing during our second night was good
($\sim$ 0.8 -- 1\arcsec) and conditions were photometric.  During our
observations the parallactic angle rotated during our observations from
PA$_{parallactic} \approx 75$\deg\ to $-$75\deg\ (hour angle $-3^h$ to
$+3^h$), and the maximum airmass was 1.45.

In order to calibrate the instrumental polarization position angle
($PA_{pol}$) correction and the polarization efficiency, we observed
the star BD+28$^{\circ}$4211 through UV and IR polaroid filters in four
waveplate positions. We find the $PA_{pol}$ correction to be very
stable (consistent with that measured on previous runs with the
polarimeter) and the polarization measurement efficiency to be $\approx
1.000\pm 0.007$; hence, no correction for the latter was made to the data. 
We also observed the zero polarization standard GD~319 as a check
on any residual polarization effects: our measurement yields $P<0.16$\%
(2$\sigma$). The electric vector position angle zero point offset was calibrated
using observations of the polarization standard stars HD204827 and
HD245310 (Schmidt, Elston, \& Lupie 1992); our measurements are in very 
good agreement with the published values. 

The data were corrected for overscan bias and flat--fielded using
internal lamps taken immediately preceding and following the
observations.  The flux calibration was performed using observations of
Feige~110 and Feige~34 (Massey \etal 1988, Massey \& Gronwall 1990)
obtained both with and without an order sorting OG570 filter in order
to correct for the second order light contamination in the spectral
region $\lambda > 7500$\AA.  All of the spectroscopic reductions were
performed using the NOAO IRAF package. The spectropolarimetric analysis
was carried out using our own software, and is based on the methods
described in Miller, Robinson, \& Goodrich (1988).

\section {Results}

\subsection {Polarimetry}

Our polarimetric results are presented in Table~\ref{poldata}.  Since
\4c\ is very faint (peak continuum $V$ surface brightness $\approx
21.2$~mag/\ebox\arcsec), the continuum polarization was determined in
large spectral bins (cf., Dey \etal 1996).  For each wavelength bin, we
tabulate the normalized linear Stokes parameters, $Q =
(I_0-I_{90})/(I_0+I_{90})$ and $U = (I_{45}-I_{135})/(I_{45}+I_{135})$,
where $I_{\theta}$ is the effective intensity of light polarized with
its electric vector at position angle $\theta$.  Since the polarization
is low, we also tabulate an unbiased estimate of the percentage
polarization $P_{unb}\equiv \pm\sqrt{\vert P^2 - \sigma_P^2\vert}$,
where $P_{unb}$ is defined to be negative if $P<\sigma_P$ (Wardle \&
Kronberg 1974, Simmons \& Stewart 1985).  The continuum emission,
sampled in these bins over the 3\farcs42$\times$1\farcs0 extracted
aperture (\ie, sampling the brightest continuum emitting region in the
galaxy), is found to be unpolarized, with formal 2$\sigma$ upper limits
of $P_{unb} < 2.4\%$ in the wavelength range $\lambda\lambda1223-1535$
(i.e., between Ly$\alpha$ and \ion{C}{4}), and $< 7\%$ in the region
$\lambda\lambda1645-1895$ (\ie, between \ion{He}{2} and \ion{C}{3}]).
Within our measurement errors, the narrow emission lines appear to be
unpolarized.  The Galactic latitude of \4c\ is 17.5\deg\ and the
interstellar percentage polarization is estimated to be low ({\it
E(B--V)}$\approx$0.15, which implies $P_{ISM}<1.4\%$; Burstein \&
Heiles 1982; Appenzeller 1968, Mathewson \& Ford 1970). A slightly
better constraint on the interstellar polarization results from the
formal 2$\sigma$ upper limit of $< 0.9\%$ for the polarization of the
narrow component of the Ly$\alpha$ emission line.

The fractional polarization of the continuum emission in \4c\ is
therefore much lower than that measured in the lower redshift ($z\sim
1-2$) radio galaxies, which are generally found to have rest-frame UV
continuum percentage polarizations of $\simgt 10\%$ (di Serego
Alighieri \etal 1989, Cimatti \etal 1996, 1997, Dey \etal 1996).  This
difference in polarization is not simply the result of the different
rest-frame wavelength ranges sampled at $z\approx 3.8$ and $z\sim 1-2$.
Spectropolarimetric observations of the $z\sim 1$ radio galaxies show
either no significant wavelength dependence of the continuum
polarization in the near UV (after accounting for dilution by
starlight) or, in a few cases, some evidence for increasing $P$ with
decreasing wavelength. Moreover, the spectropolarimetry of \4c\ may be
directly compared with existing observations of 3C~256, a $z\approx
1.82$ powerful radio galaxy for which spectropolarimetric observations
have been made down to 1400\AA\ in the rest frame (Dey \etal 1996). The
continuum polarization of 3C~256 at wavelengths shortward of \ion{C}{4}
is high ($\approx 10\%$) and $P(\lambda)$ is roughly wavelength
independent between 1400\AA\ and 2700\AA. These results suggest that,
in comparison, \4c\ has intrinsically low polarization.

\subsection {Emission Line Spectrum}

\begin{figure}
\plotfiddle{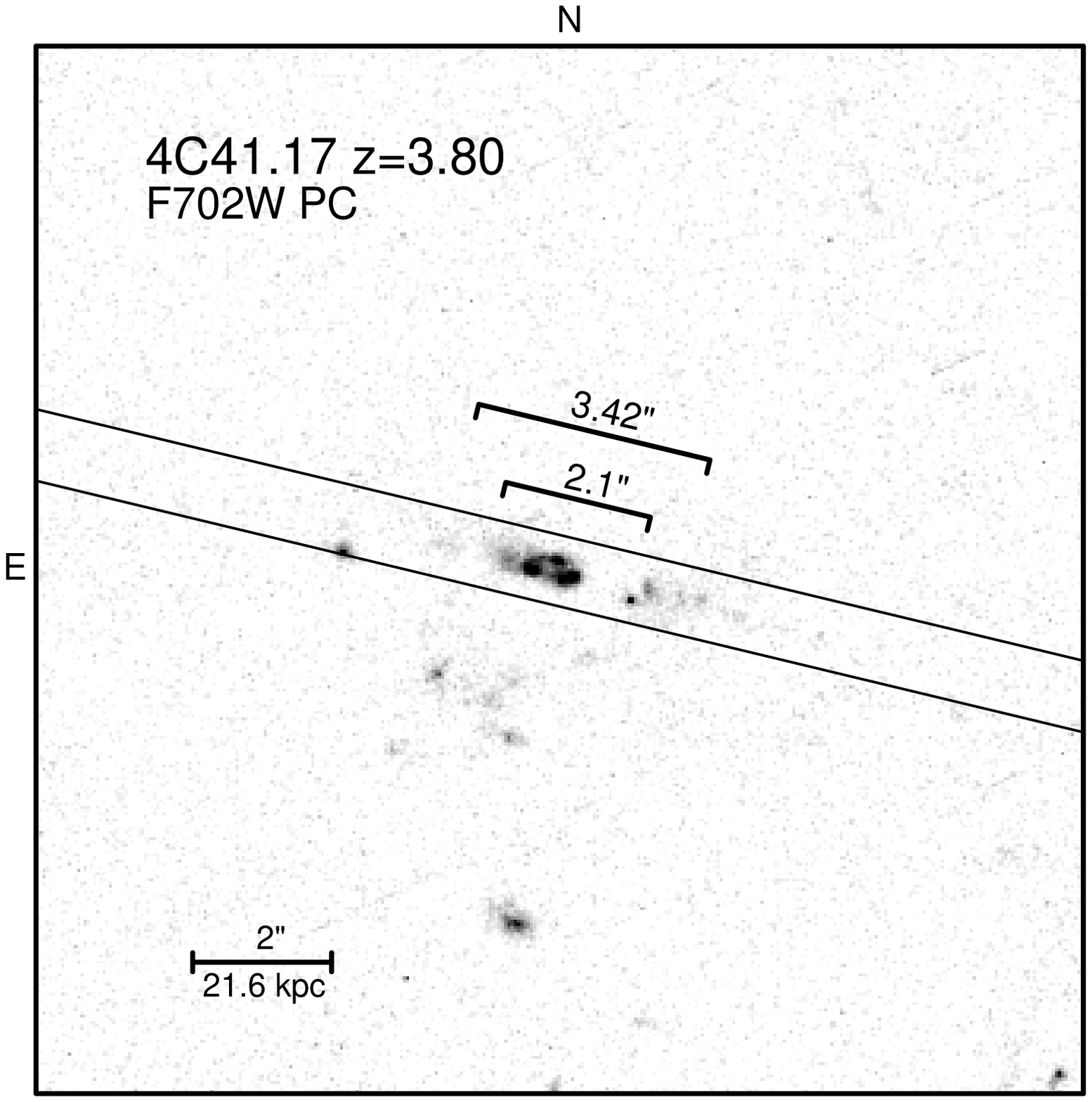}{6in}{0}{100}{100}{-330}{-150}
\figcaption{Broad band F702W image of \4c\ from van Breugel \etal 1997.
The field of view is $\approx 15\arcsec\times 15\arcsec$. The galaxy is
at $\alpha=06^h50^m52{\secper}16,\ \delta=41^\circ
30^\prime 30{\farcs}8$ (J2000). The parallel lines denote the position and
orientation of the 1\arcsec\ slit used in our Keck LRIS
spectropolarimetric observations oriented in PA=76.5\deg. The scale 
shown is for a cosmology with \hnot=50\hnotunit and \qnot=0.1. 
\label{4cf702}
}
\end{figure}

The summed spectrum of \4c\ was corrected for Galactic absorption using
the extinction curve of Cardelli \etal (1989) and an $E(B-V)$=0.15. The
parameters of the emission and absorption lines listed in
Tables~\ref{emdata} and \ref{absdata} were derived from the dereddened
spectrum (using SPECFIT in IRAF; Kriss 1994) extracted in a
$2\farcs1\times1\farcs0$ aperture centered on the brightest region of
the galaxy (Figure~\ref{4cf702}).  The redshift of the galaxy
determined from the \ion{He}{2}$\lambda$1640 emission line is
3.79786$\pm$0.00024; as expected, the resonance lines (Ly$\alpha$,
\ion{N}{5}, \ion{Si}{4} and \ion{C}{4}) result in a slightly higher
redshift, since the blue side of the lines are modified by associated
absorption. The \ion{C}{3}]$\lambda$1909 emission line may be slightly
blue shifted (by $\approx 175\pm110$\kms) relative to the systemic
velocity determined from \ion{He}{2} emission line, but since it is
also much broader (see below) than any of the other emission lines, we
consider it likely that it arises in a different region than the bulk
of the narrow--line emitting gas.  The \ion{He}{2} redshift is
consistent with (but possibly slightly redshifted by $\sim
135$\kms\ relative to) the average redshift determined from the low
ionization absorption lines.  Since most of the absorption lines may be
contaminated by (stronger) emission and have asymmetric profiles, in
this paper we adopt the \ion{He}{2} redshift as the systemic redshift
of the galaxy.

The ratios of the narrow emission line strengths of \ion{C}{3}],
\ion{C}{4} and \ion{He}{2} can be well matched by a simple
photoionization model incorporating a power--law ionizing source with
$\alpha=-1.5$ ($f_\nu \sim \nu^\alpha$), a high ionization parameter
$U\simlt 0.1$ and solar metallicity clouds (e.g., Villar-Martin,
Tadhunter \& Clark 1997). The large \ion{C}{4}/\ion{C}{3}] ratio also
favors a geometry where the illuminated sides of the line--emitting
clouds are viewed directly (Villar-Martin \etal 1996); if the
photoionizing source is the central AGN, this implies that the bulk of
the line--emitting region (the NE part of the galaxy) lies on the {\it
far} side of the AGN nucleus. This geometry is also supported by the
radio rotation measure (Carilli \etal 1994), which implies a larger
Faraday optical depth to the eastern part of the source than to the
western lobe.

\begin{figure}
\plotfiddle{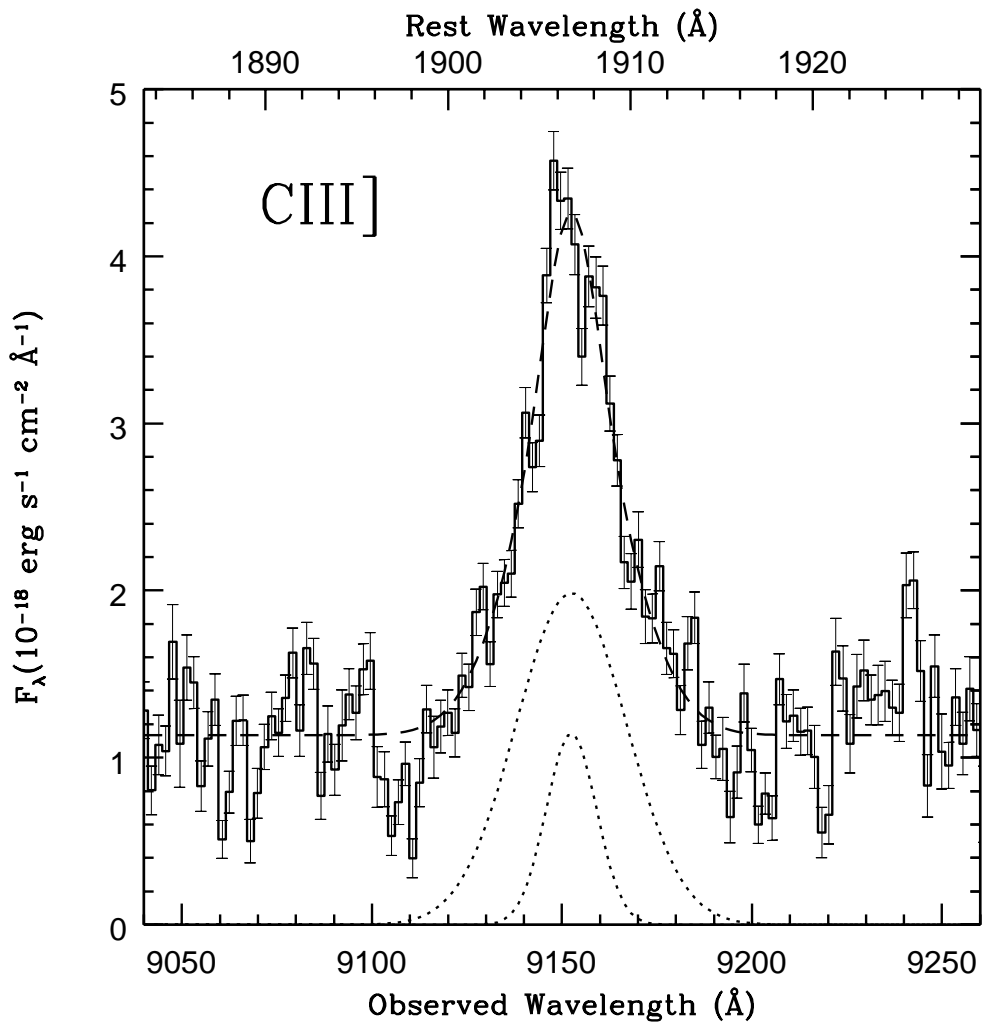}{4.5in}{0}{90}{90}{-285}{-145}
\figcaption{Two-component Gaussian fit to the CIII] emission line.  The
solid line denotes the observed spectrum measured in a
$2\arcsec\times1\arcsec$ aperture (along with 1$\sigma$ error bars),
and the dashed line the fit to the profile obtained by forcing the
broad and narrow components to have the same central wavelength.  The
dotted lines at the bottom of the plot show the broad and narrow
components of the fit.  
\label{ciii} } 
\end{figure}

The profiles of both Ly$\alpha$ and \ion{C}{3}]$\lambda$1909 reveal 
relatively broad components. On fitting these lines
with a two-Gaussian model, we find that the broad components have FWHMs
of $\approx$1100 -- 1400~\kms\ (e.g., Figure~\ref{ciii}).  Although
large, these widths may be simply a high-velocity component of the
emission line gas resulting from the interaction of the radio source
with the ambient medium, or due to entrainment of the gas in the radio
jet.  The fact that the emission lines are unpolarized supports this 
suggestion. Alternatively, the broad components may
arise in the nuclear broad-line region, and may imply that there is a
small contribution to the UV continuum and line flux from the active
nucleus.  Although the measured FWHM of the emission lines are smaller
than those typically measured in the broad line regions of steep
spectrum radio-loud quasars (${\rm FWHM_{\rm QSO} \sim 4000}$\kms, with
almost all having ${\rm FWHM_{\rm QSO}\simgt 2000}$\kms; e.g.,
Brotherton \etal 1994, Corbin 1991), the rest--frame equivalent width
of the broad component of the \ion{C}{3}] emission line ($\approx
13$\AA) is roughly comparable to that observed in these quasars. It is
therefore possible that there may be some contribution from AGN light
to the UV flux, at least at wavelengths $\lambda_{rest} \sim 1900$\AA. We note
that although we do not detect any significant polarization of the
\ion{C}{3}] line, the errors are large in this part of the spectrum;
the 2$\sigma$ upper limit from these data is large ($<11\%$), and
therefore does not provide a useful constraint.

\subsection {Absorption Line Spectrum}

\begin{figure}
\plotfiddle{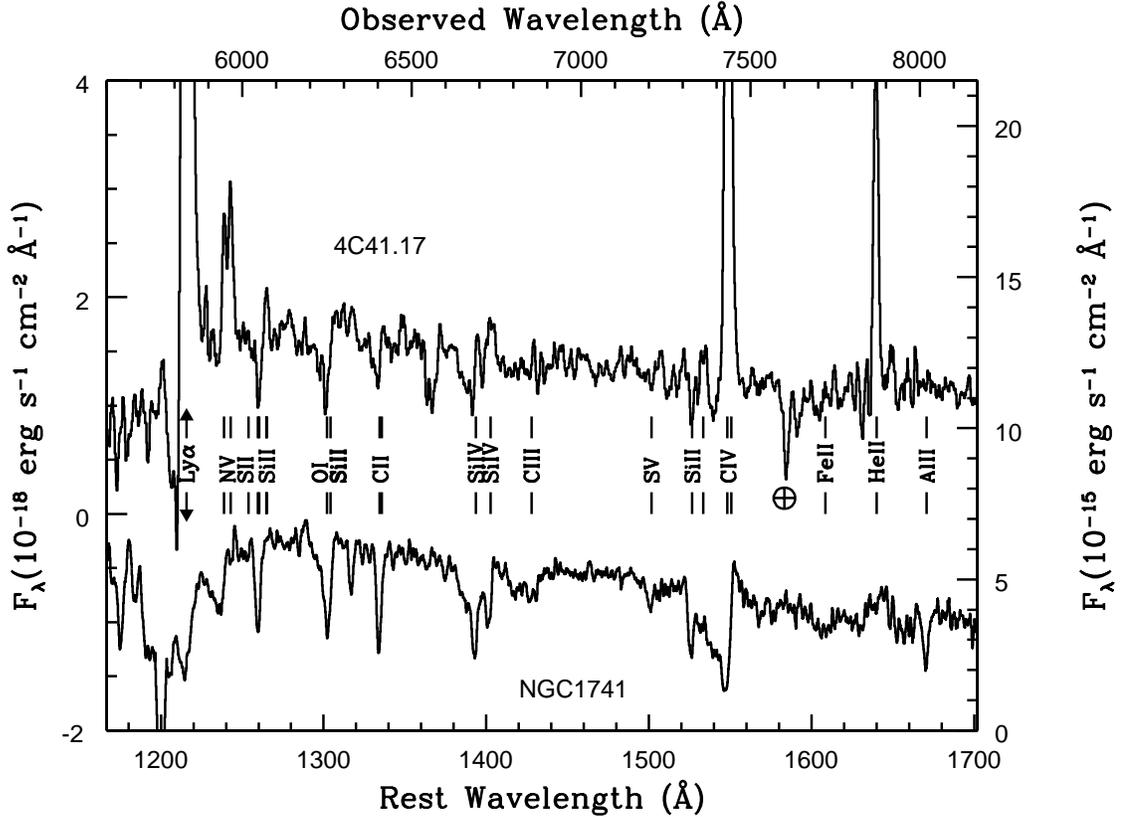}{4in}{-90}{70}{70}{-280}{380}
\figcaption{Total light spectrum of the central $2\arcsec\times
1\arcsec$ of \4c\ compared with the UV spectrum of the B1 star-forming 
knot in the nearby Wolf-Rayet starburst 
galaxy NGC~1741 from Conti et al.~(1996). The ordinate is labelled with 
the flux density scales for \4c\ and NGC~1741B1 on the left and right 
axes respectively. The two spectra show many similarities in their 
absorption line properties, although the emission line spectrum of \4c\ 
is dominated by processes related to the AGN. 
The absorption spectrum is also similar to that of the recently
discovered population of starburst galaxies at $z\simlt 3$.
\label{spectrum}
}
\end{figure}

One of the most remarkable results from our observations of \4c\ is the
detection of strong absorption features in its total light UV
spectrum.  Figure~\ref{spectrum} shows the total light spectrum of the
brightest $2\farcs1\times1\farcs0$ region of the galaxy (smoothed using
a 5-pixel boxcar), which clearly shows strong absorption lines of
\ion{Si}{4}, \ion{Si}{2}, \ion{C}{4}, \ion{C}{2}, \ion{O}{1} and
Ly$\alpha$. Some of these features have `P-Cygni--like' profiles, \ie,
an emission component juxtaposed with a blue-shifted absorption
component. P-Cygni--like profiles may arise either in stellar winds or
be the result of a different mechanism (e.g., galaxian outflows); we
discuss the possible physical origin of these profiles in \4c\ in
\S~4.2. The spectrum of \4c\ in this regard is similar to that of
actively star--forming regions in nearby galaxies such as 30~Doradus
(Vacca \etal 1995), NGC~4214 (Leitherer \etal 1996) and the B1
star-forming knot in the nearby starburst galaxy NGC~1741 (hereafter
referred to as NGC~1741B1, Conti, Vacca \& Leitherer 1996; see also
Figure~\ref{spectrum}), and to the $z\sim 2 - 3$ population of
star--forming galaxies recently discovered by virtue of their Lyman
limit absorption (e.g., Steidel \etal 1996, Giavalisco \etal 1996,
Lowenthal \etal 1997).

The parameters of the strongest detected absorption lines are listed in
Table~\ref{absdata} along with limits on the strengths of known stellar
absorption features. The absorption lines that are largely uncontaminated
by strong emission have rest--frame equivalent widths
of $\approx 1 - 2$~\AA\ 
(Table~\ref{absdata}), and are therefore slightly smaller than in most
starbursts systems (e.g., NGC~1741B1 has equivalent widths of $\approx
2$~\AA\ for these lines; Conti \etal 1996).  Many of the absorption
features have asymmetric profiles with blue wings extending to more
than 2500~\kms\ from the line centroid, and some show P-Cygni--like
profiles.  If we define the systemic velocity to be that measured from
the \ion{He}{2} emission line, then the absorption troughs
associated with Ly$\alpha$ and \ion{C}{4} are blue shifted relative to
the systemic velocity by $\approx 1700$\kms, whereas the absorption in
the low-ionization species of \ion{Si}{2}, \ion{C}{2} and \ion{O}{1}
are at the systemic velocity, or perhaps slightly blue shifted by
$\approx 135\pm 43$\kms.  The blue shifts observed in Ly$\alpha$ and
\ion{C}{4} are somewhat difficult to determine accurately since these
absorption features are contaminated by very strong line emission.

In addition to the absorption lines that are clearly associated with
P-Cygni--like emission, there are a few features for which we have only
tentative identifications. These features are likely to be either very
high-velocity blue-shifted absorption arising in a outflowing gas, or
due to foreground (intervening) metal-line systems.  The strong
absorption feature at 6555\AA\ appears to be (at the present
resolution) a composite of three absorption lines at 6544.3\AA,
6560.0\AA\ and 6573.6\AA\ (Figure~\ref{SiIV}.  The ratios of the first
two features suggest that these may be
\ion{Mg}{2}$\lambda\lambda$2796.35,2803.53 in an intervening system at
$z=1.340$. We tentatively identify the 6573.5\AA\ line with the
\ion{O}{5}$\lambda$1371 wind feature in \4c. This line is strong in the
spectra of O3 giant and supergiant stars (Walborn \etal 1995a).  An
unambiguous understanding of the true origin of this complex must await
higher resolution and higher signal-to-noise ratio data.

Table~\ref{absdata} also lists absorption lines at observed
wavelengths of 6351.5\AA, 6661.8\AA, and 7256.5\AA. These features may
correspond to the \ion{C}{2}$\lambda$1335, \ion{Si}{4}$\lambda$1400,
and \ion{Si}{2}$\lambda$1527 features blue--shifted by
$\approx$2500\kms\ from the systemic velocity. Alternatively, the
6661.8\AA\ feature is more likely to simply be the blue wing of the \ion{Si}{4}
absorption profile, and the 6351.5\AA\ line may correspond to a similar
feature at $\lambda\approx 1322$\AA\ observed in NGC~1741B1
(Figure~\ref{spectrum}).  In summary, absorption line systems
physically associated with \4c\ are observed at approximately the
systemic velocity defined by the line--emitting gas, with the
high--ionization P-Cygni--like lines extending to velocities of $\simgt
1700$\kms, and possibly also at $\approx$2500\kms\ blue shifted
relative to the line-emitting gas.

\section {Discussion}

\subsection{Scattered Light and Starlight in \4c}

The majority of radio galaxies at redshifts $z\sim 1$ exhibit UV
continuum polarization, with the position angle of the electric vector
generally perpendicular to the major axis of the UV extent (e.g.,
Tadhunter \etal 1988, di Serego Alighieri \etal  1989, Jannuzi \&
Elston 1991, Jannuzi \etal 1995, Dey \etal 1996, Cimatti \etal 1996,
1997). In almost all known cases, the percentage polarization is
large ($\simgt 10\%$) at rest-frame UV wavelengths and, in a
few cases, monotonically decreases with increasing wavelength. In a few
well--studied cases, broad emission lines have been detected both in
the total light and polarized flux spectra. These results have led to
the conclusion that the spatially extended, aligned UV morphologies of
the $z\sim 1$ powerful radio galaxies are largely a result of scattered
light from the AGN.  \4c\ is therefore an exception to the rule --- the
low measured percentage polarization does not support the hypothesis
that the UV emission is dominated by light from the AGN.

However, the lack of polarization does not necessarily {\it prove} that
the UV continuum emission from \4c\ does not have a scattered AGN
contribution.  The polarization could be diluted by geometrical effects
resulting from averaging over large areas (our spectral extraction
covers an aperture of 3\farcs2$\times$1\farcs0) which have different
polarization angles.  In order to investigate this possibility, we
measured the polarization in the UV continuum between Ly$\alpha$ and
CIV as a function of spatial position along the slit from our best
seeing data. Within the errors, there is no convincing detection of
polarization in the extended continuum. It is also possible to dilute
the fractional polarization by multiple scattering.  This would require
that the scattering be due to dust particles rather than electrons,
which would in turn result in large reddening and extinction of
the spectrum; no significant extinction or reddening is observed.
The upper limit on the estimated extinction ($E(B-V)\simlt 0.1$) corresponds
to a dust optical depth at a rest wavelength of 
1500\AA\ of $\tau_{\rm 1500\AA}\simlt 0.9$, which 
should not result in significant dilution of $P$ by multiple scattering.  
Finally, some theoretical models suggest that the dust scattering
efficiency may drop sharply at wavelengths less than $\lambda \sim 2200
- 2600$\AA\ (A.\ Laor, personal communication). Although such a feature
has never been observed in the scattered light UV spectra of Galactic
reflection nebulae (e.g., Calzetti \etal 1995), if this were indeed the
case, our limits may be consistent with dust scattering contributing
significantly to the longer wavelength radiation $\lambda \simgt
2600$\AA, but very little to the shorter wavelength radiation.  We
therefore conclude that the UV spectrum, 
at least at wavelengths $\lambda_{rest} \sim
1500$\AA, is largely uncontaminated by scattered AGN light.

Is it possible that other AGN-related processes dominate the UV
continuum emission from \4c?  Due to the strong Ly$\alpha$ emission
observed in \4c, it has been suggested by Dickson \etal (1995) that the
UV continuum emission in \4c\ may be dominated by thermal continuum
emission. The blue ($F_\lambda\propto\lambda^{-1.8}$) UV continuum
spectrum of \4c\ indicates that it is unlikely that the recombination
continuum emission from hydrogen dominates the continuum emission in
the region between Ly$\alpha$ and \ion{C}{4}.  The flux in the
Ly$\alpha$ emission line can provide a fairly good estimate of the
contribution of the nebular emission to the UV spectrum in our
2\farcs1$\times$1\farcs0 aperture. However, this resonance line is
easily attenuated by dust, and instead a more reliable estimate may be
derived from the \ion{He}{2}$\lambda$1640 emission line. For a solar
metallicity gas ionized by a power-law continuum (ionization parameter
${\rm log}U=-1.8$ to $-2.8$), the typical ratio of
Ly$\alpha$/\ion{He}{2} is $\approx 20$. Hence, assuming that
$F_{total}({\rm Ly\alpha}) \approx 1.1\times 10^{-15}~{\rm
erg\ s^{-1}\ cm^{-2}}$, and that the typical temperature of the
line-emitting gas is $10^4$~K, the total contribution of recombination
and bremsstrahlung emission from \ion{H}{1} at $\lambda < 2600$\AA\ is
$<1\times 10^{-19}~{\rm erg\ s^{-1}\ cm^{-2}\ \AA^{-1}}$, or $<7$\% of
the UV continuum emission (e.g., Osterbrock 1989, Brown \& Mathews
1970). Hence, it is unlikely that any obvious AGN-related process (\ie,
scattered light, thermal bremsstrahlung and recombination emission)
contributes significantly to the UV continuum emission from \4c, at
least at rest wavelengths $\lambda_{rest}\sim 1500$\AA.  We are
therefore left with the possibility that this UV continuum emission is
dominated by starlight.

It is possible, however, that the longer wavelength radiation has a
larger contribution from AGN emission. For wavelengths longward of the
\ion{He}{2}$\lambda$1640 line, the strong telluric OH emission
precludes a high signal-to-noise measurement of the fractional
polarization, and the upper limits from our measurements are less
restrictive (e.g., $P_{2\sigma} < 7\%$ for $1645 < \lambda_{rest} <
1885$\AA). In addition, the \ion{C}{3}] emission line is observed to
have a relatively broad component with a rest--frame equivalent width similar to
that observed in quasars, and the redshift determined from the line is
slightly different from that determined from the \ion{He}{2} narrow
line which suggests that this line arises in a different region (at a
different velocity) from the bulk of the narrow--line emitting gas.  It
is therefore possible that there is a significant AGN contribution to
the emission at the longest observed wavelengths. Indeed, the
observed continuum emission from the central body of the galaxy can be
fairly well--modelled by a double power--law: the region between
Ly$\alpha$ and \ion{C}{4} has a $F_\lambda\propto \lambda^{-1.8}$
dependence, whereas the spectral region between \ion{He}{2} and
\ion{C}{3}] is roughly wavelength independent.  The hypothesis of an
AGN contribution rising to redder wavelengths would imply that the AGN
is intrinsically much redder ($\beta > +1.7$, where $F_\lambda^{AGN}
\propto \lambda^{\beta}$) than that observed in the luminous,
steep--spectrum radio--loud quasars (typically $\beta\sim -1$). In
light of the submillimeter detection of warm dust emission from
\4c (Dunlop \etal 1994), 
it is possible that its active nucleus is still enshrouded in dust
and heavily reddened at least along our line of sight.  If the polarization in
the region $1645 < \lambda_{rest} < 1885$\AA\ is real, it is noteworthy
that the electric vector position angle is parallel to the radio axis,
rather than perpendicular as it is in most $z\sim 1$ radio galaxies;
this result suggests a different scattering geometry or polarization
mechanism.  Spectroscopic and polarimetric observations at near-IR
wavelengths are necessary to determine the AGN contribution, if any, to
the observed emission at longer wavelengths.

\subsection{The Origin of the Absorption Lines}

It is of great importance to understand the origin of the absorption
lines in a galaxy observed at such an early epoch (for
\hnot=50\hnotunit, \qnot=0.1, $\Lambda$=0, the Universe is only 2.2~Gyr
old at $z$=3.8). As discussed in the previous section, the UV continuum
emission at $\lambda_{rest}\sim 1500$\AA\ cannot be adequately
explained by AGN-related processes, and is therefore likely to be
dominated by starlight.  Strong stellar UV continuum emission can be
produced by very young populations, in which the integrated UV spectrum
is dominated by the flux from hot, high-mass stars. Hence, it is
tempting to interpret the UV absorption lines observed in \4c\ as
photospheric or dense wind features from hot, young stars.  However, it
is important to note that most of the strong resonance lines detected
near the systemic velocity in the UV spectra of nearby star-forming
galaxies are generally dominated by interstellar rather than stellar
components (\eg, York et al.\ 1991).  In this section we investigate
whether or not the absorption line spectrum of \4c\ contains a
component due to starlight, and the possibility that the absorption
spectrum can arise in a galaxian outflow.

\begin{figure}
\plotfiddle{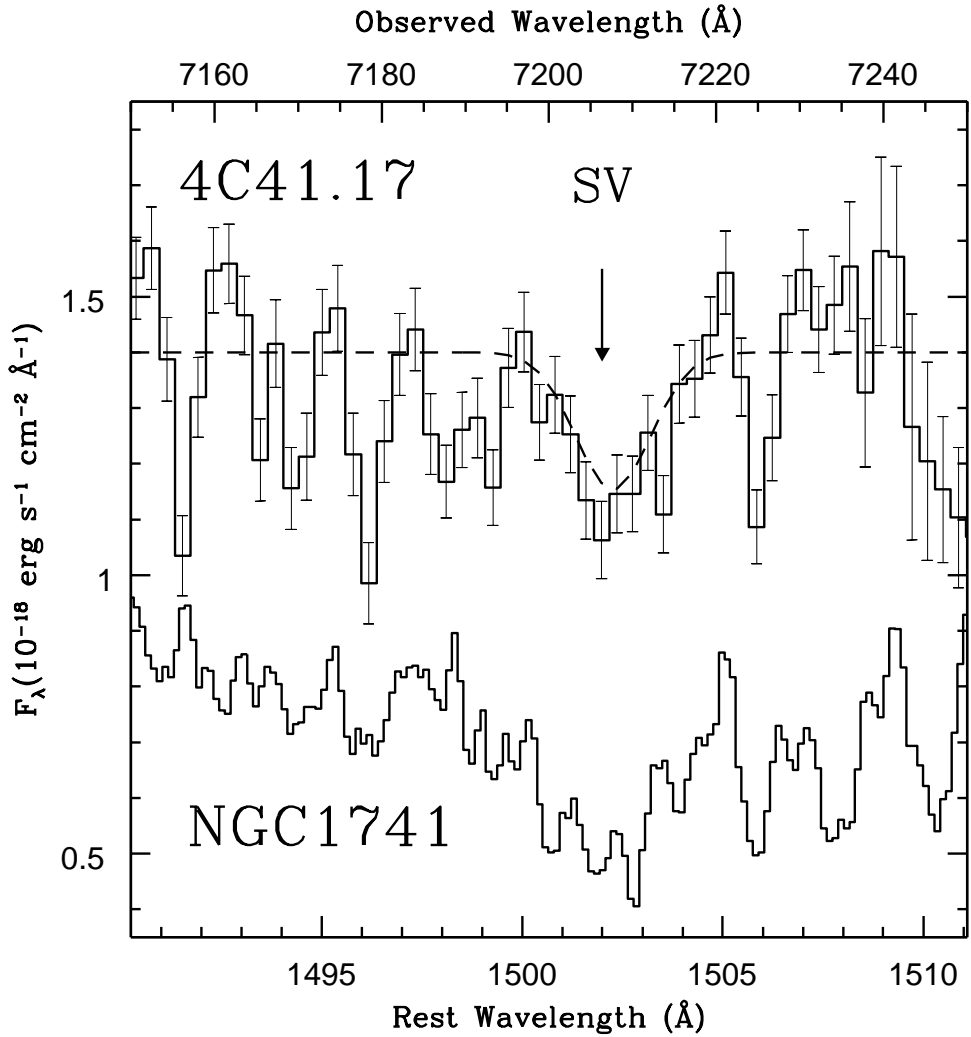}{4.5in}{0}{90}{90}{-285}{-145} 
\figcaption{The unsmoothed total light spectrum of the central $2\arcsec\times
1\arcsec$ of \4c\ compared with the {\it HST} spectrum of NGC~1741B1
(Conti \etal 1996) in the spectral region surrounding the
SV$\lambda$1502 stellar photospheric absorption line. Formal 1-$\sigma$
error bars are plotted for the spectrum of \4c. The spectrum of
NGC~1741B1 has been scaled to the match the continuum level of the
4C~41.17 spectrum, and then offset by 0.7 units. A gaussian fit to the
SV absorption line feature is shown. 
\label{SV} 
} 
\end{figure}

A clear test of whether or not the UV continuum emission is due to
young stars is the detection of stellar photospheric lines which are
uncontaminated by interstellar absorption in the integrated spectrum
(e.g., \ion{C}{3}$\lambda$1247, \ion{Si}{3}$\lambda$1296,
\ion{Si}{3}$\lambda$1417, \ion{S}{5}$\lambda$1502,
\ion{N}{4}$\lambda$1720; Leitherer, Robert \& Heckman 1995, Heckman \&
Leitherer 1997, Heckman \etal 1997; M.\ Pettini personal
communication).  These lines arise from excited levels (e.g., the lower
energy level of the \ion{S}{5}$\lambda$1502 transition is the 3$p$(J=1)
level which lies $\approx$15.8eV above ground; Bashkin \& Stoner 1975).
While these features can be easily formed in the dense, hot winds of
early-type stars, they are unlikely to arise in the cooler, less dense
interstellar medium.  The integrated spectrum of \4c\ shows at least
one unambiguous signature of starlight:  we detect the weak
\ion{S}{5}$\lambda$1502 photospheric absorption line\footnote{The
identification of the stellar absorption feature at 1502\AA\ is
uncertain at present; Dean and Bruhweiler (1985) suggest the feature is
due to \ion{Mn}{5} whereas Willis \etal (1986) identify absorption and
nearby emission at 1500-1505\AA\ in the spectra of WN stars as due to
\ion{P}{3} and \ion{S}{5}. Howarth (1987) attributes the
1501.8\AA\ feature observed in the UV spectrum of the sdO star
HD~128220 to \ion{S}{5}.  In all cases, the line arises from an excited
level and is therefore photospheric. It is observed in the spectra of O
stars (Walborn \etal 1995a) and gradually weakens with increasing
spectral type; it disappears by mid-B spectral types. Given the
relative cosmological abundances of S, P and Mn and their ionization
potentials, it is most likely that this feature is \ion{S}{5}.}
(Figure~\ref{SV}), with an equivalent width of $W_\lambda({\rm
SV})\approx 0.4$\AA\ in the rest frame. The strength of the \ion{S}{5}
line in \4c\ is comparable to that observed in nearby starbursts
($W_\lambda({\rm SV}) \approx 0.5$\AA\ in NGC~1741B1). This suggests
that AGN light does not contribute significantly to the spectrum at
these wavelengths, although the signal-to-noise ratio and resolution of
the present data do not provide a strong constraint on this issue.  We
also marginally detect weak absorption due to stellar photospheric
lines of \ion{Si}{3}$\lambda\lambda$1294.6,1296.7 ($W_\lambda({\rm
SiIII}) \approx 3.7$\AA).

\begin{figure}
\plotfiddle{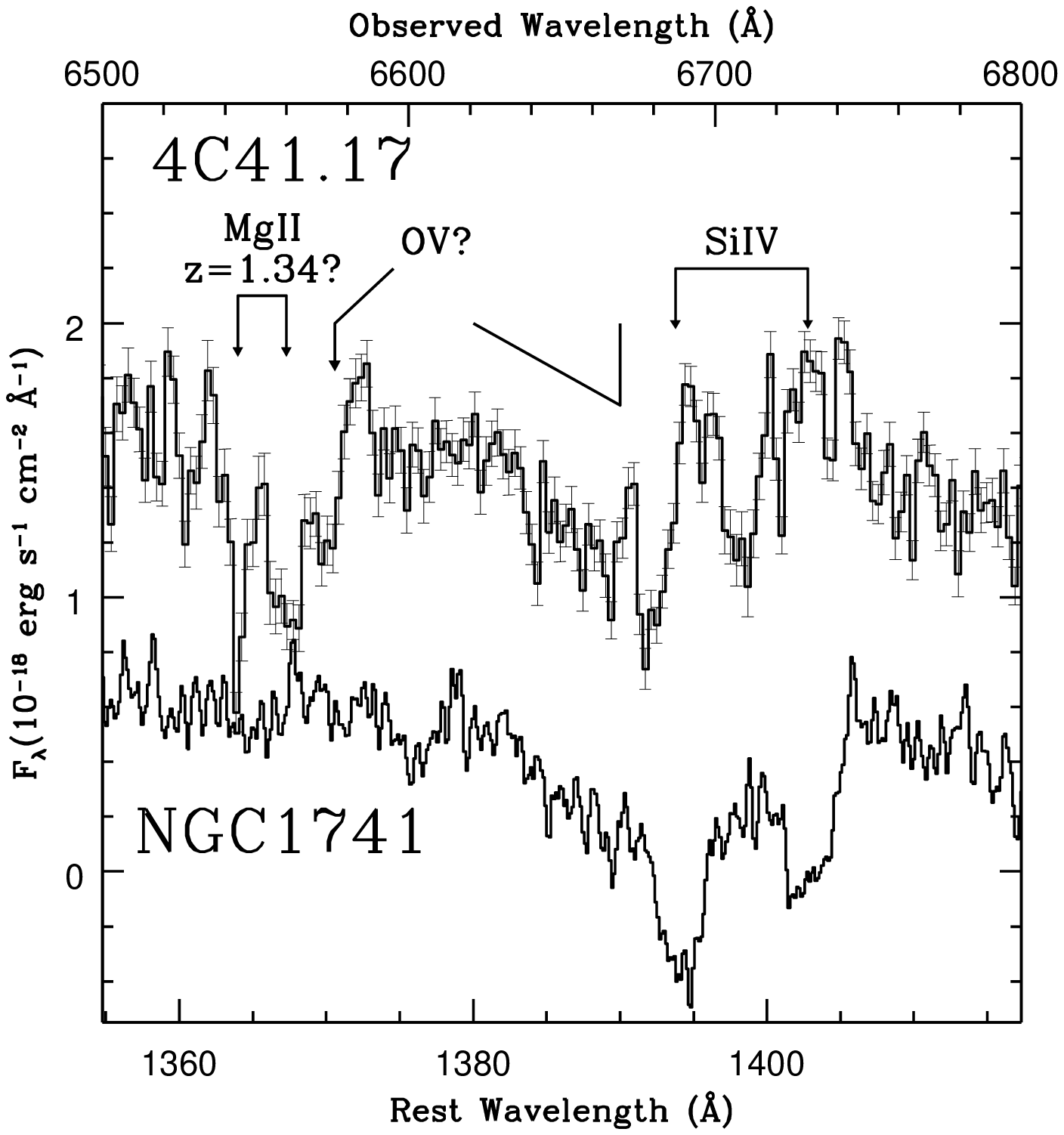}{4.5in}{0}{90}{90}{-285}{-165} 
\figcaption{The unsmoothed total light spectrum of the central
$2\arcsec\times 1\arcsec$ of \4c\ compared with the {\it HST} spectrum
of NGC~1741B1 (Conti \etal 1996) in the spectral region of the
SiIV$\lambda\lambda$1393.8,1402.8 absorption lines. Formal 1-$\sigma$
error bars are plotted for the spectrum of \4c. The spectrum of
NGC~1741B1 has been scaled to the match the continuum level of the
4C~41.17 spectrum, and then offset by 1.1 units. The connected arrows
at $\lambda_{rest}\sim1400$\AA\ represent the systemic velocity of the
SiIV lines as determined from the HeII emission redshift. Note that the
narrow absorption components at the systemic velocity are largely
filled in by emission.  The narrow components immediately blueward of
the arrows are likely to be largely interstellar. However, the blue
wing (denoted by the wedge and also observed in the spectrum of
NGC~1741B1) may be formed in winds from hot stars. The two blue
components of the composite absorption feature at 6555\AA\ are probably
due to a foreground MgII absorption system at $z\approx 1.34$. The
reddest component may have a contribution from the OV$\lambda$1371 O
star wind feature.
\label{SiIV} 
} 
\end{figure}

The shape of the \ion{Si}{4} absorption profile also indicates that hot
stars make a significant contribution to the UV spectrum.
Figure~\ref{SiIV} shows that the \ion{Si}{4} feature has narrow
components that appear slightly blue shifted from the systemic
velocity as well as a blue wing that extends to more than
2500\kms\ from the systemic velocity. Although the narrow components
are almost certainly interstellar, the shape and extent of the blue
wing closely resembles that observed in NGC~1741B1 (Figure~\ref{SiIV};
Conti \etal 1996), where it is believed to arise in hot stellar winds
(e.g., Conti \etal 1996, Heckman \& Leitherer 1997; in particular, see
their Figure 2).  Finally, as reported in \S~3.3, there is a marginal
detection of the \ion{O}{5}$\lambda$1371 wind feature
(Figure~\ref{SiIV}). If the existence of this feature can be verified
with higher resolution data (to separate the \ion{O}{5} line from the 
foreground \ion{Mg}{2} absorber), it will provide strong support for the
presence of early spectral type O giants and supergiants (Walborn \etal
1995a).  We have also searched for other pure photospheric absorption
lines (e.g., \ion{C}{3}$\lambda$1247 and \ion{N}{4}$\lambda$1720), but
these are undetected in the present data.  Unfortunately, the upper
limits on the strengths of the undetected lines do not provide any
useful constraints or rule out the presence of starlight in the
spectrum (see Table~\ref{absdata}).

Is it possible for the P-Cygni--like absorption lines to be formed in a
galaxian outflow rather than in a stellar wind?  Some low-ionization
lines such as \ion{Si}{2} appear to show asymmetric absorption profiles
with blue wings in \4c. This is a property rarely seen in the spectra
of starburst regions or individual hot stars (with the possible
exception of some B supergiants). Since these absorption lines are
likely to have significant interstellar components, it is certainly
possible that these low-ionization lines arise in a galaxian-scale
outflow. However, it is unlikely that the blue wing of the \ion{Si}{4}
absorption line is also formed in a dense, hot wind from the galaxy
driven by supernovae or the AGN. Supernovae--driven `super winds'
observed in some starburst galaxies generally exhibit LINER--like line
ratios and rarely show strong high-ionization lines (e.g., Lehnert \&
Heckman 1996) and are therefore unlikely to be the origin of the
features observed in \4c. Absorption observed in broad absorption line
(BAL) QSOs exhibit high-ionization lines, but these tend to show
absorption extending over a very large velocity range ($>$3000\kms) and
are not spatially extended sources.  Nuclear BAL-like absorption seen
reflected off dust and gas in the ambient medium can be ruled out by
the lack of detectable polarization in \4c.  

Nevertheless, it is worth investigating whether a massive galaxian outflow 
could 
result in the observed \ion{Si}{4} profile. A consideration of the
energetics of such an outflow and the mass of outflowing material
necessary to produce the observed high-ionization line profiles may
provide a method by which to discriminate between the `galaxian outflow'
and `stellar wind' hypotheses described above. Although a detailed
estimate of these quantities is beyond the scope of this paper, a very
rough calculation can be used to show that the amount of mass implied
by the absorption in the blue wing of \ion{Si}{4}$\lambda$1393 is 
large if this blue wing arises in an outflowing galaxian wind. The
blue wing of the \ion{Si}{4} line has a rest frame equivalent width of
$\approx 1.5$\AA. Although this line may be saturated,
the
{\it minimum} column density required to produce the line is
$N_{min}({\rm SiIV}) \approx 1.1\times 10^{14}\, (W_{\rm SiIV}/1{\rm
\AA})\, {\rm cm^{-2}}$. The mass of the \ion{Si}{4} absorbing region
(\ie, the region of the outflow with velocities between 700\kms\ and
2500\kms\ blueward of the systemic velocity) is
$$
M_{min}({\rm SiIV\,\, region}) > 8 \times 10^8 
\left({{X_{\rm Si}}\over{X_{\rm Si,cosmic}}}\right)^{-1}
d_{\rm Si}^{-1}\,
h_{50}^{-2}\, \Msun,
$$
where $X_{\rm Si}/X_{\rm Si,cosmic}$ is the abundance of Si in
\4c\ relative to the cosmic abundance, and $d_{\rm Si}$ is the
depletion of Si onto dust grains. We have assumed that the continuum
emitting region is $21.6\times 10.8$~kpc in size, and used the measured
\ion{Si}{4} rest equivalent width of 1.5\AA. For the purposes of this
order of magnitude estimate, we have assumed that the ion fraction of
\ion{Si}{4} is the maximum possible in thermal equilibrium (\ie, at
$T\sim 10^5$~K; Shull \& Van Steenberg 1982). The energy input required
to drive this minimum mass is equivalent to that of $>8\times 10^6$
supernovae. Recent determinations of the abundances in $z\sim 4$ damped
Ly$\alpha$ absorbers have resulted in estimates of ${\rm [Si/H] \approx
-2}$ (\ie, $X_{\rm Si}/X_{\rm Si,cosmic}\approx 10^{-2}$) in undepleted
gas (Lu \etal 1996), suggesting that the minimum mass estimate derived
above may be higher by at least two orders of magnitude.  Also,
$M_{min}$ corresponds solely to the mass in the region in velocity
space in which the blue wing of the \ion{Si}{4} feature is formed;
accounting for the probable line saturation, the range of ionization
states and velocities that would be present in a real outflow,
depletion of Si onto dust grains, and the integration over all solid
angles will increase this number further by at least an order of
magnitude. This estimate for the mass of the outflowing material in
front of the continuum emitting region of the galaxy is therefore
potentially very large ($\sim 10^{11}$~\Msun) and is comparable to the
{\it total} mass of material in the entire system (e.g., van Ojik \etal
1997, Chambers \etal 1990).  Although more detailed modelling of the
ionization state, velocity structure and geometry of the outflow are
required to investigate these constraints further, the
order-of-magnitude estimate presented here suggests that it is unlikely
that the blue wing of \ion{Si}{4} arises in outflowing material.  On
the other hand, photospheric emission from hot stars provides not only
the one viable explanation for the UV continuum emission observed from
\4c (\S~4.1), but also a natural origin for the blue wings of the high
ionization absorption line profiles.

One of the main differences between the spectrum of \4c\ and those of
nearby star--forming galaxies is that the spectrum of \4c\ is dominated
by strong line emission, which is likely the result of photoionization
by the hard spectrum of the active nucleus and possibly shock
ionization resulting from the interaction of the radio source plasma
with the ambient medium. As discussed in \S3.2 the
\ion{C}{3}]/\ion{C}{4}/\ion{He}{2} ratios suggest a high ionization
parameter with high energy photons, implying that the observed emission
lines are not due to photoionization by starlight, but instead to
photoionization by the hard AGN spectrum (e.g., Villar--Martin \etal
1996) or fast shocks (e.g., Dopita \& Sutherland 1996).  The AGN-driven
emission lines dilute the stellar and interstellar absorption lines
weakening their observed strengths, and overwhelm the much weaker
emission component that may be associated with stellar wind features
(\ie, the `true' P-Cygni profiles).  Although the emission lines are
not powered by hot stars, it is intriguing that a fairly good fit to
the overall spectrum of \4c\ can be produced by adding strong emission
lines to the spectrum of NGC~1741B1, suggesting that the UV {\it
continuum} emission shortward of \ion{C}{4} may indeed be starlight.

In summary, although the present data are inadequate to unambiguously
discriminate between stellar wind and galaxian outflow origins for the
resonance absorption transitions, several lines of argument suggest
that the UV continuum emission at wavelengths $\simlt 1600$\AA\ is
dominated by light from a young, hot stellar population. The detection
of \ion{S}{5}, the marginal detection of \ion{Si}{3}, the blue wing of
the \ion{Si}{4} profile, the unpolarized continuum emission, the
inability of AGN-related processes to account for the UV continuum flux
and the overall similarity of the spectra of 4C~41.17 and NGC~1741B1
together provide a compelling argument for the existence of starlight
in this $z=3.8$ galaxy. If the stellar populations in \4c\ and
NGC~1741B1 have similar metallicities and initial mass funtion (IMF),
the similarity of the shape and equivalent width of the blue wing of
\ion{Si}{4} in \4c\ and NGC~1741B1 may reflect comparable numbers of
hot stars per unit mass in the two systems. The high-ionization
resonance lines may therefore have both stellar and interstellar
components: our observations are consistent with the hypothesis that
the high-velocity blue-wings of the \ion{Si}{4} and \ion{C}{4}
absorption are produced in fast winds from hot stars, whereas the
absorption associated with the lower ionization states arises in the
larger scale dense, galaxian wind. A test of this hypothesis can be
provided by detections of the fainter stellar photospheric lines, but
this must await higher signal-to-noise and higher resolution
spectroscopy. The discussion in the remainder of this paper is based on 
our favoured interpretation that the UV continuum spectrum of \4c\ is 
dominated by the emission from a young stellar population. 

\subsection{The Age of the Stellar Population} 

If the observed UV continuum emission is indeed due entirely to
starlight, the slope of the continuum and the profiles of the stellar
absorption lines can be used to constrain the global properties of the
stellar population. Since our spectral extraction samples a large
volume of the galaxy (our 2\farcs1$\times$1\farcs0 aperture
corresponds to 21.6$\times$10.8~$h_{50}^{-1}$~kpc for \qnot=0.1), it is
very plausible that we are observing a mix of stellar populations
corresponding to a range of ages rather than a burst of a single age.
Nevertheless, the absorption lines in the integrated spectrum can
provide a few constraints on the luminosity-weighted mean age of the
stellar population dominating the UV light. Detailed spectral synthesis
of the UV spectrum of \4c\ will be presented by Vacca \& Dey (1997); in
this section, we present some preliminary conclusions based on the
general appearance of the spectrum.

We may constrain the upper mass cutoff ($M_{\rm up}$) of the IMF
and the age of the stellar population by using the
\ion{Si}{4} absorption feature.  As discussed above, although the
strong, narrow low-velocity components of the \ion{Si}{4} doublet are
largely interstellar, the high-velocity blue wing of the profile
(Figure~\ref{SiIV}) is probably formed in hot stellar winds.  The
\ion{Si}{4} line is a powerful age diagnostic, as it only appears in
the integrated light of young populations (ages $<$1~Gyr), reaching
maximum strength in B0/B1 stars (Leitherer et al.~1996), with broad,
blue-shifted wings of the profile being formed in supergiant winds.
Since this absorption line is observed only in massive stars, the very
presence of a broad \ion{Si}{4} line implies an $M_{\rm up}>40$\Msun.
In addition to the \ion{Si}{4} line, the \ion{S}{5}$\lambda$1502
absorption line is weakly detected in the integrated spectrum. This
line is only observed in the spectrum of O-type stars, is weaker in
late O and early B subtypes, and no longer detectable in the spectra of
mid B subtypes (Walborn \etal 1985, 1995b).  Hence, the UV light is
dominated by a very young population, implying that the stars were
formed in a burst less than a few million years before, or (more
likely) that we are observing \4c\ during an epoch of ongoing star
formation.

An additional constraint is provided by the shape of the UV spectrum.
The population synthesis models of Leitherer and Heckman (1995)
demonstrate that the UV spectrum of a young star-forming population can
never be bluer than $F_\lambda\propto \lambda^{-2.6}$, and should
gradually redden with age. Hence, the observed slope of the spectrum of
\4c\ ($\lambda^{-1.8}$) provides an upper limit on the age and the
extinction.  In the absence of any extinction, the constant
star-formation solar metallicity Salpeter IMF 
models of Leitherer \& Heckman (1995)
imply maximum ages for the stellar population of $\simlt$~600~Myr.  If
there is a moderate amount of extinction then the intrinsic slope of
the UV spectrum of \4c\ will be bluer than the observed slope and the
age will be younger.  The extinction for \4c\ is probably not
larger than $E(B-V) \sim 0.1$ mag, because the intrinsic slope would
then be bluer than the bluest model values. (Here we have assumed the
extinction law given by Kinney et al.\ 1994. The change in the slope as
a function of reddening for this extinction law is given by Leitherer
et al.\ 1996.)

For the instantaneous burst model, the UV continuum slope should become
redder much faster than in a continuous star formation scenario.  Using
the same arguments about the continuum slope as those given above, we
can place an upper limit of about 16~Myr for the age of the
population.  For an instantaneous burst model, the line profiles place
much stronger constraints on the age and $M_{up}$ than in the
continuous (constant-rate) star formation case, since high-velocity blue wings
are only observed for ages less than 6~Myr and
$M_{up}>30$\Msun. If the age is less than 6~Myr, then the intrinsic
spectral shape of the continuum emission must be between
$F_\lambda\propto \lambda^{-2.1}$ and $F_\lambda\propto\lambda^{-2.6}$,
which implies that the observed continuum is moderately reddened ({\it
E(B--V)}=0.04 to 0.10). Again, this age estimate is based on the
assumption that the \ion{Si}{4} emission component is from hot stars
rather than from the AGN.

An additional constraint on the stellar content of \4c\ may be derived
from existing near-infrared photometry. The near-infrared $J$ and
$K_{\rm S}$ bands are largely uncontaminated by strong line emission
from \4c.  In the 1\farcs0$\times$2\farcs1 spectroscopic aperture
discussed here, \4c\ has measured magnitudes of $J=21.8\pm0.4$ and
$K_{\rm S}=20.7\pm0.5$ (Adam Stanford, personal communication). Under
the null hypothesis that all the observed near-infrared and optical
emission is starlight, we find that unreddened, young continuous
(constant-rate) star-forming models with ages less than $\sim300$~Myr
and instantaneous burst models with ages less than $\sim 16$~Myr that
match the observed optical spectrum do not fit the observed
near-infrared fluxes. In fact, these simple unreddened models do not
provide satisfactory fits to both the $J$ and $K_{\rm S}$ fluxes at any
age, either overpredicting the $K_{\rm S}$ flux or underpredicting the
observed $J$ flux. In contrast, slightly reddened ($E(B-V)=0.1$),
models with ages 4--20~Myr (for instantaneous burst models) or
10~--300~Myr (for continuous star-forming models) provide a
satisfactory fit to both the overall optical-IR spectral energy
distribution and (at the young age end) to the \ion{Si}{4} line
profile. If the observed infrared emission contains a reddened AGN
component, the spectral energy distribution of the underlying
population is bluer than observed, and therefore younger. If the
population is indeed young ($\simlt$1~Gyr), then a reddened AGN
can contribute no more than 60-65\% of the total $K_{\rm S}$ flux.

In summary, if the galaxy is forming stars continuously, the age
(derived from its UV spectrum) is less than 600~Myr.  The age estimates
presented in this section are at present phenomenological and are
uncertain as they depend upon several assumptions regarding the shape
of the IMF, the star-formation history, the metallicity of the
star-forming population, and the reddening intrinsic and along the
line-of-sight to \4c. Nevertheless, it is of considerable interest that
a simple model can provide an adequate fit to both the observed optical
and near-infrared continuum fluxes of \4c, suggesting that stronger
constraints may be eventually derived when high signal-to-noise
moderate resolution optical and infrared spectra of the UV light and the
4000\AA\ or Balmer-break region are available.

\subsection{The Structure of the Galaxy and the Case for Triggered Star--formation}

In a companion paper, van Breugel \etal~(1997) present deep {\it HST}
imaging of \4c, and demonstrate that the brightest central region is
composed of several compact ($\simlt 5$~kpc) knots, which are
distributed in an `edge-brightened' structure suggestively lying on the
outer envelope of the central radio source structure (also see
Figure~\ref{4cf702}). The deep spectroscopic data presented in this
paper strongly suggest that the UV absorption--line spectra of these
knots are similar to that of star--forming regions in nearby galaxies.

If the compact UV knots in \4c\ are indeed star--forming regions, then
the spatial distribution of these knots relative to the radio source
emission is suggestive of models where star-formation may be triggered
by the radio source (De Young 1989, Begelman \& Cioffi 1989).  De Young
(1989) proposed that as radio jets propagate into dense protogalactic
gas, star--formation occurs in the compressed post--shock gas as it
cools. This results in a spatial distribution of hot stars that is both
aligned with the radio source and edge--brightened in the direction
normal to the radio jet axis, on a scale of roughly $1-4$~kpc from the
jet axis depending upon the jet parameters and age of the stellar
population (De Young 1989).  
In the scenario proposed by Begelman \& Cioffi (1989),
star--formation also occurs when intergalactic clouds are overtaken and
compressed by the expanding over-pressured cocoon surrounding the
radio source. This mechanism can result in a large star--formation rate
($\sim 100$\Msun/yr; Begelman \& Cioffi 1989) and also results in young
stars in an edge--brightened structure (De Young 1989). 
The observed distribution of
star--forming clumps in \4c\ is edge--brightened as predicted in these
scenarios, and the spatial extent in the transverse direction is indeed
comparable to that expected. The strongly asymmetric radio source
morphology (shorter eastern arm) and brightness distribution (the
eastern knots are more luminous both in the radio and the optical) also
support a picture of stronger ongoing interactions between the radio
source and the ambient medium on the eastern side of the galaxy.

Although the evidence presented here (spectroscopy coupled with the van
Breugel \etal 1997 {\it HST} imaging data) is circumstantial, at
present there is no clearer support for this picture than these data.
Our observations of \4c\ are consistent with the interpretation that
the alignment effect in at least this $z=3.80$ radio galaxy is due to
star formation induced by the radio source, and not due to scattering
as is the case in the bulk of the $z\sim 1$ radio galaxies.

If the continuum emission is indeed due to starlight from a young
population, we can use the observed UV continuum luminosity to estimate
the star-formation rate in the system. The specific luminosity at
$\lambda_{rest}\approx1500$\AA\ is $L_{1500} \approx 2.0 \times
10^{42}h_{50}^{-2}\ {\rm erg\ s^{-1}\ \AA^{-1}}$ (\qnot=0.1). For a
Salpeter IMF with an upper mass cutoff of 80\Msun,
the stellar population synthesis models of Leitherer \etal (1993)
suggest that a continuously star--forming population has ${\rm
log}~L_{1500} = 39.25 - 40.16$ for ages between 1 and 9~Myr after the
onset of star-formation.  Hence, if {\it all} of the 1500\AA\ flux in
\4c\ is due to young stars, the star--formation rate is between 140 --
1100~$h_{50}^{-2}$~\Msun/yr. If we account for screen extinction of
$E(B-V)$=0.1, these estimates are increased by a factor of
$\approx$2.9.  Although this range is large, at the lower end, this
rate is observed for extreme starburst galaxies.  For a high--redshift
galaxy, possibly in its early stages of formation, such a large rate
may be expected. At this sustained rate of star--formation, it will
only take $< 0.7$~Gyr to form a $10^{11}$~\Msun\ system of stars, a timescale
comparable to the dynamical timescale. This
rate of star--formation is higher than that observed in the $z\sim 2-3$
``Lyman--limit'' galaxies discovered by Steidel \etal (1996), which are
typically observed to have star--formation rates at $z\sim 3$ of
$10-70~h_{50}^{-2}$~\Msun/yr (\qnot=0.1). The lower rates observed in
these systems may be due to the fact that these galaxies are being
observed at a later stage in their history ($\Delta t\approx
0.75h_{50}^{-1}$~Gyr between $z=4$ and $z=3$ for $\Omega=0.2$), or that
these are lower--mass systems than \4c, or that their star-formation 
rates have been under-estimated due to dust-extinction. 

The timescale derived above for the age of the starburst can be
compared with the expansion timescale of the radio source. In the
instantaneous burst model, the age of the stellar population is less
than 16~Myr. During this time, the radio source will expand by roughly
16.4~$(v_{\rm exp}/1000\kms)(\tau/16{\rm Myr})$~kpc, or $\sim
1\farcs5~h_{50}(v_{\rm exp}/1000\kms)(\tau/16{\rm Myr}){\rm sin}\theta$
from the location at which it triggered the star formation, where
$v_{\rm exp}$ is the radio source expansion speed, $\tau$ is the age of
the population, and $\theta$ is the angle between the direction of
expansion and the line of sight.  The lack of a bright core component
to the radio source and the absence of any significant AGN component to
the UV emission suggest that the radio source is oriented close to the
plane of the sky, with our direct view of the nuclear region obscured by 
dust.  Hence, since the UV continuum emission is observed
to be roughly coincident with the radio emission in the central regions
(Miley et al.~1992, Carilli \etal 1994), we may conclude that the
triggering was recent, or that the radio source is expanding slowly, or
both.

The large star--formation rate derived for \4c\ requires a large
reservoir of cold gas. Indeed, there is some evidence for an extended
cold component from the observation of spatially extended Ly$\alpha$
absorption in \4c\ (Hippelein \& Meisenheimer 1993, van Ojik \etal
1997, Dey \etal 1997$b$) suggesting at least $\sim 4\times
10^7h_{50}^{-2}$\Msun\ of cold gas in the central regions, and $\sim
10^9h_{50}^{-2}$\Msun\ in a large extended halo (\qnot=0.1).  In addition, the
detection of submillimeter emission from \4c\ implies that the
galaxy contains a very large mass in dust ($M_{\rm dust}\sim 8\times
10^8h_{50}^{-2}$\Msun\ for \qnot=0.1; Dunlop \etal 1994), and suggests
the presence of such a reservoir. Under the assumption that the dust is
heated purely by starlight, Dunlop \etal derive star--formation rates
of $\sim 5\times 10^3h_{50}^{-2}$\Msun/yr (\qnot=0.1); although there
may be a significant AGN contribution to the dust heating, it is
noteworthy that this estimate is of the same order of magnitude as the
estimates derived above from the UV luminosity alone. The blue
UV/optical color and lack of strong color gradients in \4c\ implies
that the observed continuum emission is not very reddened, and is
likely to be indicative of an asymmetric distribution of dust. We
speculate that \4c\ may well be a young galaxy emerging from its dust
cocoon, with only the very outer regions visible at present.

\subsection{Is \4c\ a Typical Evolutionary Predecessor of the $z\sim1$
Powerful Radio Galaxies?}

In addition to its different polarization properties, \4c\ also
exhibits an important morphological difference at rest-frame UV and
optical wavelengths compared to lower redshift objects.  The $z\sim 1$
3CR radio galaxies tend to be clumpy and spatially extended in the
rest-frame UV, whereas their rest-frame optical morphologies (observed
near-IR) are generally more symmetric; \ie, rather than reflecting the
peculiar UV structures, the optical structures look more like `normal'
elliptical galaxies.  \4c\ also has a clumpy UV morphology (van Breugel
\etal 1997), but its rest-frame optical morphology is also clumpy and
extended along the major axis of the radio emission (Chambers \etal
1990); in fact, there is near spatial coincidence between the
rest--frame UV and optical emission (Graham \etal 1995). Although the
rest wavelengths sampled at these different redshifts are 
different, the $z\sim 1$ radio galaxies exhibit a
noticeable morphological change across the 4000\AA\ break (i.e., the
approximate wavelength beyond which stars begin to dominate over the AGN
component), whereas \4c\ ($z=3.80$) does not exhibit this
property.

The dissimilarity between the polarization and morphological properties
of \4c\ and those of the lower redshift powerful radio galaxies raises
the intriguing question of whether we are witnessing an evolutionary
phenomenon. A tempting hypothesis is that radio galaxies at $z\simgt 4$
are in the process of formation: they are forming stars on the dense
borders of the expanding cocoon inflated by the radio lobes (e.g.,
Begelman \& Cioffi 1989), and their UV and optical emission is
dominated by the light from young stellar populations. The AGN
component, scattered or directly viewed, is likely to be diluted by the
young starlight. In contrast, in the $z\sim1$ radio galaxies, the
population has aged and dynamically relaxed, and no longer contributes
any significant flux to the UV spectrum. The UV light in the $z\sim1$
population may instead be dominated by other blue components, such as
the AGN light (viewed directly or scattered by dust and electrons in
the ambient medium) remnant star--forming regions, or Balmer continuum
emission from the photo-- and shock--ionized gas.  In an \hnot=50,
$\Omega$=0.2 Universe, the time between $z=3.8$ and $z=1$ is $\approx
4.8$~Gyr, providing ample time for the population to age and (possibly)
dynamically relax to explain the colors and morphology of the stellar
component in $z\sim 1$ radio galaxies. Recent spectral synthesis
modelling of the UV spectra of $z\sim 1.5$ lower-power radio galaxies
which have no discernible light contribution from the AGN result in age
estimates of $\simgt$~3.5~Gyr, consistent with this picture
(Dunlop \etal 1996, Spinrad \etal 1997, Dey \etal 1997$a$).

\section{Conclusions}

We have presented deep spectropolarimetric observations of one of the
most distant known radio galaxies, \4c\ at $z=3.80$. The spectrum is
dominated by strong emission lines (possibly excited by shocks and / or
photoionization by the AGN), a bright UV continuum emission and fairly
strong resonance absorption lines of both low-- and high--ionization
species. The UV continuum emission from the brightest regions in this
galaxy is unpolarized ($P_{2\sigma}<2.4\%$), suggesting that scattered
AGN light may not be a dominant contributor to the UV flux from the
galaxy. Many of the lines exhibit P-Cygni--like profiles, \ie, emission
components associated with blue-shifted absorption components. The UV
absorption spectrum is similar to that of star--forming regions in
nearby starburst galaxies, and to the spectra of the
recently-discovered high-redshift population of starburst systems. Most
of the strong low-velocity narrow absorption lines most likely arise in
the interstellar medium of \4c. However, we report a detection of the
\ion{S}{5}$\lambda$1502 absorption line which arises in the
photospheres of O-type stars and is uncontaminated by interstellar
absorption, and also a possible detection of the \ion{Si}{3} stellar
photospheric lines. We also find that the shape of the high-velocity
blue wing of the \ion{Si}{4} doublet is similar to that observed in
nearby starburst galaxies such as NGC~1741, where it is interpreted as
arising in an O star wind. It appears unlikely that the blue wing of
the \ion{Si}{4} line arises in a galaxian outflow since the implied
mass in such an outflow is very large.  These considerations, combined
with the lack of detectable polarization, the lack of strong AGN-like
broad-line emission, the inability of other AGN-related processes to
produce the bulk of the continuum emission, and the similarity of
{\4c}'s spectrum to that of the star-forming region NGC~1741B1,
together provide a compelling argument that the UV spectrum of \4c\ is
dominated by starlight from a young stellar population.

If the UV continuum emission is entirely dominated by starlight,
instantaneous burst solar metallicity population synthesis models imply
an age of less than 16~Myr, and suggest that the spectrum is reddened
by dust either intrinsic to \4c\ or along the line of sight. If the
star-formation is continuous, then the age of the population is $\simlt
600$~Myr.  Under the assumption that {\it all} of the 1500\AA\ flux in
\4c\ is due to young stars, the star--formation rate is between 140 --
1100~$h_{50}^{-2}$~\Msun/yr.  A fairly young, slightly reddened
population provides an adequate fit to the overall spectral energy
distribution, but there potentially could be a red AGN which dominates
the near-infrared emission.

\4c\ is one of the highest redshift aligned radio galaxies known. Our
data suggest that the UV properties of this galaxy are different from
those of $z\sim1$, aligned radio galaxies: the unpolarized UV continuum
emission and the edge--brightened, clumpy {\it HST} morphology combined
with the spectroscopic evidence of a starburst-like spectrum may imply
that the UV light is dominated by young stars forming on the dense edge
of the pressurized cocoon formed by the radio source, as suggested by
Begelman \& Cioffi (1989).  If so, we are witnessing this galaxy in the
process of formation.  The young age derived for the starburst in \4c\ is
consistent with the observation that the radio continuum emission is
roughly spatially coincident with the UV emitting regions. The ages
determined for the $z\sim 1$ radio galaxies are consistent with the
hypothesis that these objects form the bulk of their stars at $z\sim 4$
in a low-density Universe.

\acknowledgements

We thank Ron Quick, Randy Campbell, Tom Bida, Bob Goodrich and Mike
Brotherton for invaluable help during our Keck run. We are grateful to
Claus Leitherer for providing us with the spectrum of NGC~1741B1, and
to Dave De Young, Mike Brotherton, Mike Dopita, Buell Jannuzi, Joan
Najita and Adam Stanford for useful discussions and comments on our
manuscript.  We thank Adam Stanford and Daniel Stern for sharing their
near-infrared photometry of \4c\ in advance of publication. We thank
the anonymous referee who provided useful suggestions.  The
W.\ M.\ Keck Observatory is a scientific partnership between the
University of California and the California Institute of Technology,
made possible by the generous gift of the W.\ M.\ Keck Foundation.
R.\ A.\ acknowledges support from NSF grant AST 93-21441.  Part of this
research was performed at IGPP/LLNL under the auspices of the
U.\ S.\ Dept.\ of Energy under contract \# W-7405-ENG-48.

\pagebreak

\centerline {\bf References}

\medskip


\refindent Appenzeller, I.\ 1968, \apj, 151, 907.

\refindent Barthel, P.\ D.\ 1989, \apj, 336, 606. 

\refindent Bashkin, S.\ \& Stoner, J.\ O.\ 1975, ``Atomic Energy
Levels and Grotrian Diagrams'' (North Holland Publishing 
Co.: Amsterdam), vol.~2

\refindent Begelman, M.\ C.\ \& Cioffi, D.\ F.\ 1989, \apj, 345, L21

\refindent Brotherton, M.\ S., Wills, B.\ J., Steidel, C.\ C., \&
Sargent, W.\ L.\ W. 1994, \apj, 423, 131.

\refindent Brown, R.\ L.\ \& Mathews, W.\ G.\ 1970, \apj, 160, 939
 
\refindent Burstein, D.\ \& Heiles, C.\ 1982, \aj, 87, 1165

\refindent Calzetti, D., Bohlin, R.\ C., Gordon, K.\ D., Witt, A.\ N.\ \& 
Bianchi, L.\ 1995, \apj, 446, L97.

\refindent Cardelli, J., Clayton, \& Mathis, 1989, \apj, 345, 245

\refindent Carilli, C.\ L., Owen, F.\ N.\ \& Harris, D.\ E.\ 1994, \aj, 107, 480

\refindent Chambers, K.C., Miley, G.K. \& van Breugel, W. 1987, \nature, 329, 604.

\refindent Chambers, K.\ C., Miley, G.\ K.\ \& van Breugel, W.\ J.\ M.\ 1990, 
\apj, 363, 21.

\refindent Cimatti, A., Dey, A., van Breugel, W., Antonucci, R.\ \& Spinrad, 
H.\ 1996, \apj, 465, 145. 

\refindent Cimatti, A., Dey, A., van Breugel, W., Hurt, T.\ \& Antonucci, 
R.\ 1997, \apj, 476, 677

\refindent Conti, P.\ S., Leitherer, C.\ \& Vacca, W.\ D.\ 1996, \apj, 461, L87

\refindent Corbin, M.\ R.\ 1991, \apj, 375, 503

\refindent Dean, C.\ A.\ \& Bruhweiler, F.\ C.\ 1985, \apjsupp, 57, 133

\refindent Dey, A.\ \& Spinrad, H.\ 1996, \apj, 459, 133

\refindent Dey, A., Cimatti, A., van Breugel, W., Antonucci, R.\ \& 
Spinrad, H.\ 1996, \apj, 465, 157

\refindent Dey, A., Spinrad, H., Stern, D., Dunlop, J., Peacock, J., 
Jimenez, R.\ \& Windhorst, R.\ 1997$a$, in preparation

\refindent Dey, A., van Breugel, W., Antonucci, R.\ \& Spinrad, H.\ 1997$b$, 
in preparation.

\refindent De Young, D.\ S.\ 1981, \nature, 293, 43.

\refindent De Young, D.\ S. 1989, \apj, 342, L59

\refindent Dickson, R., Tadhunter, C., Shaw, M., Clark, N.\ \& Morganti, R.\ 1995, 
\mnras, 273, L29

\refindent di Serego Alighieri, S., Binette, L., Courvoisier, T.\ J.--L., 
Fosbury, R.\ A.\ E., \& Tadhunter, C.\ N.\ 1988, \nature, 334, 591.

\refindent di Serego Alighieri, S., Cimatti, A., \& Fosbury, R.\ A.\ E. 1993, \apj, 404, 584.

\refindent di Serego Alighieri, S., Cimatti, A., \& Fosbury, R.\ A.\ E. 1994, \apj, 431, 123.

\refindent di Serego Alighieri, S., Fosbury, R.\ A.\ E., Quinn, P.\ J.\  
\& Tadhunter, C.\ N.\ 1989, \nature, 341, 307.

\refindent Dopita, M.\ A.\ \& Sutherland, R.\ S.\ 1996, \apjsupp, 102, 161

\refindent Dreher, J.\ W., Carilli, C.\ L., Perley, R.\ A.\ 1987, \apj, 316, 611

\refindent Dunlop, J.\ S., Hughes, D.\ H., Rawlings, S., Eales, S.\ A.\ \& 
Ward, M.\ J.\ 1994, \nature, 370, 347

\refindent Dunlop, J.\ S., Peacock, J., Spinrad, H., Dey, A., Jimenez, R., 
Stern, D.\ \& Windhorst, R.\ 1996, \nature, 381, 13

\refindent Eales, S.\ A.\ \& Rawlings, S.\ 1993, \apj, 411, 67

\refindent Giavalisco, M., Steidel, C.\ C.\ \& Macchetto, F.\ D.\ 
1996, \apj, 470, 189

\refindent Goodrich, R.\ W., Cohen, M.\ H.\ \& Putney, A.\ 1995, \pasp, 107, 179. 

\refindent Graham, J.\ R.\ \etal 1994, \apj, 420, L5

\refindent Heckman, T.\ M., Gonzalez-Delgado, R., Leitherer, C.,
Meurer, G.\ R., Krolik, J., Wilson, A.\ S., Koratkar, A.\ \& Kinney,
A.\ 1997, \apj, 482, 114.

\refindent Heckman, T.\ M.\ \& Leitherer, C.\ 1997, \aj, in press.
 
\refindent Hippelein, H.\ \& Meisenheimer, K.\ 1993, \nature, 362, 224

\refindent Howarth, I.\ D.\ 1987, \mnras, 226, 249

\refindent Jannuzi, B.\ T.\ \& Elston, R. 1991, \apj, 366, L69.

\refindent Jannuzi, B.\ T., Elston, R., Schmidt, G., Smith, P., \&
Stockman, H.\ 1995, ApJ, 454, L111.

\refindent Kinney, A.\ L., Calzetti, D., Bica, E.\ \& Storchi-Bergmann,
T.\ 1994, \apj, 429, 172

\refindent Kriss, G.\ A.\ 1994, in Proceedings of the 3rd Conference on
Astrophysics Data Analysis \& Software Systems, ASP Conf.\ Ser.\ v. 61,
ed.\ D.\ R.\ Crabtree, R.\ J.\ Hanisch, \& J.\ Barnes.

\refindent Lehnert, M.\ D.\ \& Heckman, T.\ M.\ 1996, \apj, 462, 651

\refindent Leitherer, C.~\& Heckman, T.\ M.\ 1995, \apjsupp, 96, 9.

\refindent Leitherer, C., Robert, C., \& Heckman, T.\ M.\ 1995, \apjsupp, 99, 173.

\refindent Leitherer, C., Vacca, W.\ D., Conti, P.\ S., Filippenko, A.\ 
V., Robert, C.\ \& Sargent, W.\ L.\ W.\ 1996, \apj, 465, 717  

\refindent Lowenthal, J.\ D., Koo, D.\ C., Guzman, R., Gallego, J.,
Phillips, A.\ C., Faber, S.\ M., Illingworth, G.\ D.\ \& Gronwall,
C.\ 1997, \apj, in press

\refindent Lu, L., Sargent, W.\ L.\ W., Barlow, T.\ A., Churchill, C.\ W.\ 
\& Vogt, S.\ S.\ 1996, \apjsupp, 107, 475

\refindent Massey, P.\ \& Gronwall, C.\ 1990, \apj, 358, 344.

\refindent Massey, P., Strobel, K., Barnes, J.V. \& Anderson, E. 1988, \apj, 328, 315.

\refindent Mathewson, D.\ S.\ \& Ford, V.\ L.\ 1970, MemRAS, 74, 139. 

\refindent McCarthy, P.J. 1993, \araa, 31, 639

\refindent McCarthy, P.J., van Breugel, W.J.M., Spinrad, H. \& Djorgovski, S.
1987, \apjlett, 321, L29

\refindent McNamara, B.\ R., Jannuzi, B.\ T., Elston, R., Sarazin, C.\ L.\ 
\& Wise, M.\ 1996, \apj, 469, 66

\refindent Miley, G.\ K., Chambers, K.\ C., van Breugel, W.\ \& Macchetto, F.\ 1992, \apj, 401, L69

\refindent Miller, J.\ S., Robinson, L.\ B.\ \& Goodrich, R.\ W.\ 1988, in 
{\it Instrumentation for Ground-Based Optical Astronomy}, ed.\ L.\ B.\ Robinson,
(Springer-Verlag: New York), p.~157.

\refindent Oke, J.\ B., Cohen, J.\ G., Carr, M., Cromer, J., Dingizian, A., 
Harris, F.\ H., Labrecque, S., Lucino, R., Schaal, W., Epps, H., \& Miller, 
J.\ 1995, \pasp, 107, 375.

\refindent Osterbrock, D.\ E.\ 1989, Astrophysics of Gaseous Nebulae and Active 
Galactic Nuclei (University Science Books: California)

\refindent Rigler, M.\ A., Stockton, A., Lilly, S.\ J., Hammer, F.\ \& 
Le F\`evre, O.\ 1992, \apj, 385, 61

\refindent Rees, M.\ J. 1989, \mn, 239, 1P.

\refindent Schmidt, G.\ D., Elston, R.\ \& Lupie, O.\ L.\ 1992, \aj, 1563.

\refindent Shull, J.\ M.\ \& Van Steenberg, M.\ 1982, \apjsupp, 48, 95

\refindent Simmons, J.\ F.\ L.\ \& Stewart, B.\ G.\ 1985, \aa, 142, 100.

\refindent Spinrad, H., Dey, A., Stern, D., Dunlop, J., Peacock, J., 
Jimenez, R.\ \& Windhorst, R.\ 1997, \apj, in press

\refindent Steidel, C.\ C., Giavalisco, M., Dickinson, M.\ \& Adelberger, 
K.\ L. 1996, \aj, 112, 352

\refindent Steidel, C.\ C., Giavalisco, M., Pettini, M., Dickinson, M.\ \& 
Adelberger, K.\ L. 1996, \apj, 462, L17

\refindent Tadhunter C.\ N., Fosbury, R.\ A.\ E., \& di Serego
Alighieri, S. 1988, in Proc. of the Como Conference, BL Lac Objects,
ed. L. Maraschi, T.\ Maccacaro \& M.\ H.\ Ulrich (Berlin:
Springer-Verlag), 79.

\refindent Vacca, W.\ D., Robert, C., Leitherer, C.\ \& Conti, P.\ S.\ 1995, 
444, 647

\refindent Vacca, W.\ D. \& Dey, A.\ 1997, in preparation. 

\refindent van Breugel, W.\ J.\ M., Fillipenko, A.\ V., Heckman, T.\ M., 
\& Miley, G.\ K.\ 1985, \apj, 293, 83 

\refindent van Breugel, W.\ J.\ M.\ \& Dey, A.\ 1993, \apj, 414, 563

\refindent van Breugel, W.\ \etal 1997, in preparation

\refindent van Ojik, R., R\"ottgering, H..\ J.\ A., Miley, G.\ K.\ \&
Hunstead, R.\ W.\ 1997, \aa, 317, 358

\refindent Villar--Martin, M., Binette, L.\ \& Fosbury,
R.\ A.\ E.\ 1996, \aa, 312, 751

\refindent Villar--Martin, M., Tadhunter, C.\ N.\ \& Clark,
N.\ E.\ 1997, \aa, in press.

\refindent Walborn, N.\ R., Nichols-Bohlin, J., \& Panek, R.\ J. 1985,
IUE Atlas of O-Type Spectra from 1200 to 1900\AA, NASA Reference
Publication 1155

\refindent Walborn, N.\ R., Lennon, D.\ J., Haser, S.\ M., Kudritzki, R.-P., 
\& Voels, S.\ A.\ 1995a, \pasp, 107, 104

\refindent Walborn, N.\ R., Parker, J.\ Wm., \& Nichols, J.\ S. 1995b,
IUE Atlas of B-Type Spectra from 1200 to 1900\AA, NASA Reference
Publication 1363

\refindent Wardle, J.\ F.\ C.\ \& Kronberg, P.\ P.\ 1974, \apj, 194, 249.

\refindent Willis, A.\ J., van der Hucht, K.\ A., Conti, P.\ S., \& 
Garmany, D.\ 1986, \aasupp, 63, 417

\refindent York, D.\ G., Caulet, A., Rybski, P., Gallagher, J., Blades, 
J.\ C., Morton, D.\ C.\ \& Wamsteker, W.\ 1990, \apj, 351, 412


\begin{deluxetable}{llrrcrcl}
\tablewidth{0pt}
\tablecaption{Measured Continuum and Emission Line Polarizations\tablenotemark{\dag}}
\tablehead{
\colhead{$\Delta\lambda_{obs}$(\AA)} & \colhead{$\Delta\lambda_{rest}$(\AA)} & 
\colhead{$Q$(\%)} & \colhead{$U$(\%)} & \colhead{$P$(\%)} & \colhead{$P_{unb}$(\%)} &
\colhead{$\theta$(\deg)} & \colhead{Note}
}
\startdata
%
%
5500--5800 & 1146--1208 &   3.43$\pm$3.40 &$-$0.55$\pm$3.12 & 3.48 & 0.76$\pm$3.39 & $-$4.56$\pm$27.96 & \nl
5800--5870 & 1208--1223 &$-$0.71$\pm$0.26 &$-$0.91$\pm$0.26 & 1.15 & 1.12$\pm$0.26 &  115.98$\pm$6.43  & Ly$\alpha$  \nl
5825--5845 & 1213--1218 &$-$0.29$\pm$0.22 &$-$0.14$\pm$0.23 & 0.32 & 0.23$\pm$0.22 &  102.95$\pm$19.91 & Ly$\alpha$ (narrow) \nl
5870--7370 & 1223--1535 &   0.92$\pm$0.58 &   1.02$\pm$0.56 & 1.37 & 1.25$\pm$0.57 &   24.04$\pm$11.80 & \nl
7370--7480 & 1535--1558 &   4.57$\pm$1.15 &   0.57$\pm$1.14 & 4.61 & 4.46$\pm$1.15 &    3.55$\pm$7.16  & \ion{C}{4} \nl
7480--7810 & 1558--1627 &$-$2.01$\pm$1.80 &   1.06$\pm$1.70 & 2.27 & 1.41$\pm$1.78 &   76.13$\pm$22.43 & \nl
7845--7900 & 1634--1646 &$-$4.26$\pm$2.93 &$-$3.99$\pm$2.87 & 5.84 & 5.07$\pm$2.90 &  111.59$\pm$14.24 & \ion{He}{2} \nl
7895--9050 & 1645--1885 &$-$3.65$\pm$1.57 &   2.01$\pm$1.37 & 4.17 & 3.88$\pm$1.53 &   75.59$\pm$10.49 & \nl
9100--9200 & 1896--1917 &   3.99$\pm$2.58 &$-$4.56$\pm$2.45 & 6.06 & 5.52$\pm$2.51 &$-$24.43$\pm$11.85 & \ion{C}{3}] \nl
\enddata
\tablenotetext{\dag}{Measured in an aperture of size 3\farcs4$\times$1\farcs0.}
\label{poldata}
\end{deluxetable}

\begin{deluxetable}{lccccc}
\tablewidth{0pt}
\tablecaption{Emission Line Measurements}
\tablehead{
\colhead{Line} & \colhead{$\lambda_{rest}$~(\AA)} & 
\colhead{$\lambda_{obs}$~(\AA)} & 
\colhead{$F_{obs}$\tablenotemark{\dag}} & 
\colhead{FWHM (${\rm km~s^{-1}}$)} & 
\colhead{$W_\lambda^{obs}$~(\AA)} 
}
\startdata
Ly$\alpha$(narrow) & 1215.67 & 5835.5$\pm$0.1 & 771$\pm$37& 613$\pm$13   & 548  \nl
Ly$\alpha$(broad) & 1215.67 & 5835.5$\pm$0.1 & 689$\pm$41& 1373$\pm$45  & 490  \nl
\ion{N}{5} & 1238.81 & 5946.8$\pm$0.5 & 16.9$\pm$1.8  & 628$\pm$48   & 10.6 \nl
\ion{N}{5} & 1242.80 & 5965.9$\pm$0.5 & 21.8$\pm$1.9 & 628$\pm$48   & 13.7 \nl
\ion{Si}{2}& 1264.89 & 6071.6$\pm$1.1 & 6.4$\pm$1.4  & 546$\pm$106  & 4.0  \nl
\ion{Si}{4}& 1393.76 & 6691.0$\pm$1.8 & 8.6$\pm$2.3  & 1134$\pm$172 & 6.3  \nl
\ion{Si}{4}& 1402.77 & 6734.2$\pm$1.8 & 11.5$\pm$1.5 & 1134$\pm$172 & 8.4  \nl
\ion{C}{4} & 1548.20 & 7428.9$\pm$0.2 & 74.7$\pm$1.8 & 541$\pm$14   & 58.8 \nl
\ion{C}{4} & 1550.77 & 7441.2$\pm$0.2 & 57.0$\pm$1.6 & 541$\pm$14   & 44.8 \nl
\ion{He}{2}& 1640.46 & 7870.7$\pm$0.4 & 55.3$\pm$2.8 & 553$\pm$28   & 51.2 \nl
\ion{C}{3}](narrow)& 1908.73 & 9152.5$\pm$0.6 & 19$\pm$15& 511$\pm$151  & 17.0 \nl
\ion{C}{3}](broad)& 1908.73 & 9152.5$\pm$0.6 & 72$\pm$15 & 1120$\pm$135 & 64.6 \nl
\enddata
\label{emdata}
\tablenotetext{\dag}{All fluxes are measured in an aperture of 
2{\farcs}1$\times$1{\farcs}0, and are in units of $10^{-18}\ {\rm 
erg\ s^{-1}\ cm^{-2}}$. The line fluxes have been corrected for Galactic 
reddening using $E(B-V)=0.15$ and the extinction curve of 
Cardelli, Clayton \& Mathis (1989).}
\end{deluxetable}

\begin{deluxetable}{crcll}
\tablewidth{0pt}
\tablecaption{Absorption Line Measurements}
\tablehead{
\colhead{$\lambda_{obs}$~(\AA)} & \colhead{$W_\lambda^{obs}$~(\AA)} & \colhead{FWHM (${\rm km~s^{-1}}$)} & 
\colhead{Identification} & \colhead{Comments} 
}
\startdata
5799.2$\pm$0.8 & 43.1$\pm$2.0& 2022$\pm$92 & Ly$\alpha$ & P-Cygni--like \nl
               & $<$1.5      &             & \ion{C}{3}$\lambda$1247    & Stellar \nl
6047.8$\pm$0.9 & 5.3$\pm$0.9 & 558$\pm$117 & \ion{Si}{2}$\lambda$1260.4 & \nl
6226.3$\pm$2.0 & 3.7$\pm$1.2 & 664$\pm$278 & \ion{Si}{3}$\lambda\lambda$1294.6,1296.7?  & Stellar?\nl
6246.4$\pm$1.0 & 6.8$\pm$1.3 & 639$\pm$162 & \ion{O}{1}$\lambda$1302.2+\ion{Si}{2}$\lambda$1304.4 & \nl
6351.5$\pm$11  & 1.4$\pm$1.4 & 735$\pm$968 & \ion{C}{2}$\lambda$1334.5? & \nl
6396.1$\pm$2.4 & 5.5$\pm$1.2 & 947$\pm$219 & \ion{C}{2}$\lambda$1334.5  & \nl
6544.3$\pm$0.4 & 3.1$\pm$0.7 & 218$\pm$63 & \ion{Mg}{2}$\lambda$2796.4 at z=1.34? & Foreground? \nl
6560.0$\pm$1.1 & 6.6$\pm$1.2 & 654$\pm$152 & \ion{Mg}{2}$\lambda$2803.5 at z=1.34? & Foreground? \nl
6573.6$\pm$1.0 & 1.6$\pm$0.6 & 263$\pm$78 & \ion{O}{5}$\lambda$1371? & Stellar? \nl
6661.8$\pm$3.0 & 4.2$\pm$1.5 & 931$\pm$456 & \ion{Si}{4}? & \nl
6680.5$\pm$0.7 & 5.5$\pm$1.2 & 445$\pm$69  & \ion{Si}{4}$\lambda$1393.8 & P-Cygni--like \nl
6708.0$\pm$1.3 & 2.3$\pm$0.8 & 362$\pm$92  & \ion{Si}{4}$\lambda$1402.8 & P-Cygni--like \nl
               & $<$1        &             & \ion{Si}{3}$\lambda$1417   & Stellar \nl
7207.7$\pm$0.8 & 2.0$\pm$0.5 & 416$\pm$84  & \ion{S}{5}$\lambda$1502    & Stellar \nl
7256.5$\pm$3.8 & 3.5$\pm$1.3 & 754$\pm$285 & \ion{Si}{2}$\lambda$1526.7?& \nl
7326.5$\pm$2.3 & 7.2$\pm$2.2 & 739$\pm$310 & \ion{Si}{2}$\lambda$1526.7 & \nl
7386.0$\pm$5.3 & 5.0$\pm$3.0 & 748$\pm$210 & \ion{C}{4}$\lambda$1548.2  & P-Cygni--like \nl
7398.3$\pm$5.3 & 3.7$\pm$3.0 & 748$\pm$210 & \ion{C}{4}$\lambda$1550.8  & P-Cygni--like \nl
               & $<$3        &             & \ion{N}{4}$\lambda$1720    & Stellar \nl
\enddata
\label{absdata}
\end{deluxetable}



\end{document}